\documentclass[iop,numberedappendix,twocolumn]{emulateapj}
\usepackage{natbib}

\def\na{\ref@jnl{New A}}                % New Astronomy
\def\nar{\ref@jnl{New A Rev.}}          % New Astronomy Review
\def\jcap{\ref@jnl{J. Cosmology Astropart. Phys.}}
\usepackage{paralist}
\usepackage[usenames,dvipsnames,svgnames,table]{xcolor}
\usepackage[caption=false]{subfig}
\usepackage{tikz}
\usetikzlibrary{positioning,shapes,shadows,arrows}

\newcommand{\msun}{${\rm M}_\odot$}
\newcommand{\msunh}{${\rm h}^{-1}{\rm M}_\odot$}
\newcommand{\mpch}{${\rm h}^{-1}{\rm Mpc}$}
\newcommand{\pararef}[1]{\S \ref{#1}}
\newcommand{\figref}[1]{Figure \ref{#1}}
\newcommand{\tabref}[1]{Table \ref{#1}}
\newcommand{\equaref}[1]{Eq. \ref{#1}}

\newlength{\hu}
\newlength{\vu}
\tikzstyle{codes}=[rectangle, rounded corners,line width=2pt, drop shadow,
draw=blue!50!black, fill=white, text= blue!50!black,
text width =\hu,font=\tiny,minimum height= 5\vu,
rectangle split, rectangle split parts=2]

\tikzstyle{datas}=[rectangle, rounded corners,line width=2pt, drop shadow,
draw=green!40!black,fill=white,text=green!40!black,
text width =\hu,font=\tiny
]

\tikzset{>=latex}

\newlength{\onecolfigwidth}
\newlength{\twocolfigwidth}
\newlength{\thirdtwocolfigwidth}
\begin{document}

%==============================================================

\title{ELUCID -- Exploring the Local Universe with the reConstructed Initial Density
field. II: Reconstruction diagnostics, applied to numerical
  halo catalogs }

\author{
Dylan Tweed\altaffilmark{1,*},
Xiaohu Yang\altaffilmark{1,2},
Huiyuan Wang\altaffilmark{3},
Weiguang Cui\altaffilmark{4},
Youcai Zhang\altaffilmark{5},
Shijie  Li\altaffilmark{1},
Y.P. Jing\altaffilmark{1,2},
H.J. Mo\altaffilmark{6,7}
}

\altaffiltext{1}{Center for Astronomy and Astrophysics, Shanghai Jiao
  Tong University, Shanghai 200240, China}

\altaffiltext{2}{IFSA Collaborative Innovation Center, Shanghai Jiao
  Tong University, Shanghai 200240, China}

\altaffiltext{3}{Key Laboratory for Research in Galaxies and
  Cosmology, Department of Astronomy, University of Science and
  Technology of China, Hefei, Anhui 230026, China}

\altaffiltext{4}{Departamento de F\'isica Te\'orica, M\'odulo 15,
  Facultad de Ciencias, Universidad Aut\'onoma de Madrid, E-28049
  Madrid, Spain}

\altaffiltext{5}{Shanghai Astronomical Observatory, Nandan Road 80,
  Shanghai 200030, China}

\altaffiltext{6}{Department of Astronomy, University of Massachusetts,
  Amherst MA 01003-9305}

\altaffiltext{7}{Astronomy Department and Center for Astrophysics,
  Tsinghua University, Beijing 10084, China}

\altaffiltext{*}{E-mail: \href{mailto: dtweed@sjtu.edu.cn}{dtweed@sjtu.edu.cn}}

%==============================================================

\begin{abstract}
The ELUCID project aims to build a series of
realistic cosmological simulations that reproduce the spatial and
mass distribution of the galaxies as observed in the Sloan Digital
Sky Survey (SDSS). This requires powerful reconstruction techniques to create
constrained initial conditions.
We test the reconstruction method by
applying it to several $N$-body simulations.
We use 2 medium resolution simulations from each of which  three additional constrained $N$-body simulations were produced.
We compare the resulting friend of friend catalogs by using the particle indexes as
tracers, and quantify the quality of the reconstruction by varying the
main smoothing parameter. The cross identification method we use proves to be efficient, and the results suggest
that the most massive reconstructed halos are effectively traced from
the same Lagrangian regions in the initial conditions. Preliminary
time dependence analysis indicates that high mass end halos converge
only at a redshift close to the reconstruction redshift.  This suggests
that, for earlier snapshots, only collections of progenitors may be
effectively cross-identified.
\end{abstract}


\maketitle

%==============================================================

\section{Introduction}

One of the key aspects of modern cosmology is to interpret the
distribution and properties of galaxies in the sky. The diversity of
galaxy properties is inherent to their history and environment.

The formation and evolution of galaxies is generally understood within
the $\Lambda$CDM paradigm. Matter itself is dominated by a dark
component: the Dark Matter subject to gravitational
interactions. Galaxies are understood to reside in dark matter halos,
that act as a gravitational potential wells. These follow a pattern of
hierarchical structure formation, where the smaller structures form
first and merge to build larger and larger ones. The gas itself
is bound to these structures and as they build up, baryonic processes
take place leading to the formation of stars in galaxies that further
evolve and interact within those halos.

Structure formation is thus understood as a non-linear
process, that cannot be fully described analytically. Building {these
structures requires the use of specific numerical methods: $N$-body
simulations. These codes focus on the Dark Matter component, and
successfully describe the formation of structures by implementing
gravitational interactions on large scales. Dark Matter is
described numerically by data points (otherwise
  referred as particles) that trace a mass element
  corresponding to a
volume element of the early 'homogeneous' universe.

Over the decades, such
simulations have been widely used, with little variation in term of
physics. Combined with ever decreasing limitations of computer resources
and vast improvement in term of implementations, larger volumes can
be explored with increasing resolution.

Distributions and
properties of galaxies are observed with ever increasing accuracy, but
little information concerning the dominating
component can be inferred. Nevertheless various methods exist to explore the formation of
galaxies within Dark Matter halos. The most direct one is to numerically solve
the evolution of the baryonic component on top of the Dark
Matter one. This method requires more complex implementations than
$N$-body simulations, with gas
described either as \begin{inparaenum}[(i)]
\item  numerical data points with associated density \citep[Smooth
Particles hydrodynamics: ][]{Springel2001a, Springel2005, Wadsley2004},
\item grid cells fixed in the volume (cells are refined
and unrefined as required to explore high gas density while
neglecting low density regions) \citep[nested grids or Adaptive Mesh
Refinement: ][]{Kravtsov1997, Kravtsov2003, Teyssier2002}
\item or moving cells (the gas
element is associated with a numerical point within a volume
defined from the distribution of nearby mesh points through Voronoi
tesselation) \citep{Springel2010}. \end{inparaenum}
Less computationally intensive methods involve applying
models on the scale of the halo itself where evolution can be
traced with merger trees. These Semi-Analytical Models (SAM) have
been successfully applied to halo catalogs extracted from $N$-body simulations
\citep{WhiteFrenk1991, Kauffmann1993, Somerville1999, Cole2000,
  vdBosch2002, Hatton2003, Kang2005, Croton2006, Baugh2006}.

Both types of methods are used to create mock galaxy catalogs. These
catalogs are usually confronted with the observed ones and have been
quite successful in reproducing statistical features; such as the 2 point correlation
functions, luminosity functions, color distributions and star formation
rates. Both sub-grid models (applied to gas elements) and semi-analytical
models (applied to halos) require
determination of
a large number of parameters which can lead to some
degeneracies.

Even more simplified approaches are halo occupation distribution
models, where observed galaxies are assigned to halos by matching
the halo mass and stellar mass functions
\citep{Jing1998, Jing2002, Berlind2002, Bullock2002, Scranton2002,
  Yang2003, vdBosch2003, Behroozi2010,
  Rodriguez2015}. Such methods have been successful in
defining the Stellar Mass Halo Mass (SMHM) relation but the
scatter of this relation still remains uncertain and difficult to interpret.
Higher constraints can be obtained by
comparing specific galaxies, their environment and the Inter Galactic
Medium (IGM).
To achieve this, one needs to obtain one or several simulations
that match the distribution of galaxies as seen in the sky. In order
to understand how an $N$-body simulation can reproduce a given
distribution, one must have a basic understanding of how a cosmological
$N$-body code works. Basically the distribution of matter (here dark
matter only), is represented by numerical data points (particles)
describing the mass and volume element of this matter. Since dark matter
answers to gravity alone, the motion of this dark matter element obeys
the Poisson equation. A cosmological $N$-body code is essentially just a Poisson
solver, with various optimizations rendering the computation
possible. An $N$-body simulation starts at early redshift ($\sim 100$)
from a uniform distribution of particles following a
  grid. But the disposition of particles cannot be perfectly uniform
  in position and velocity since then the distribution would remain unchanged. Obtaining a realistic
simulation implies applying small shifts in both positions and velocities
on the initial grid nodes. This displacement field corresponds to the Initial
Conditions (IC). They are usually randomly generated
from the initial power spectrum for a given cosmology. From
the same set of ICs any $N$-body code would produce the
same density field. However, the formation and evolution of large
scale structures is a non-linear process. The only way to derive the
matter distribution is thus to run the numerical simulation.

Reconstructing a particular density field with an $N$-body simulation
  consists in finding, among all possible
sets of ICs, the one that would produce the closest match.
 To achieve this purpose, it is not realistic to create an
   infinite number of trial sets of ICs, run an $N$-body simulation
   from each set and chose the optimal one. The solution is to use other
numerical techniques to reverse engineer the displacement
  field (that characterize a set of ICs) from the target density
  field.

Since the pioneering work of \citet{HoffmanRibak1991} and
\citet{NusserDekel1992}, many methods have been developed to
reconstruct ICs of the local universe. These methods use either redshift surveys of galaxies \citep{HessKitauraGottloeber2013, WangMoYang2016} or radial peculiar velocities \citep{Kravtsov2002, Klypin2003, Gottloeber2010, Sorce2016} of low redshift galaxies as tracers of
the cosmic density field at the present day. In order to derive ICs from
these low redshift tracers several dynamical models are adopted in the literature;
\begin{inparaenum}[(i)]
\item linear theory \citep{HoffmanRibak1991, Gottloeber2010},
\item Lagrangian perturbation theory \citep{NusserDekel1992, Brenier2003, Lavaux2010, JascheWandelt2013, Kitaura2013, WangMoYang2013, Doumler2013} or
\item particle-mesh (PM) dynamics \citep{WangMoYang2014}.
\end{inparaenum}
These dynamical models can be
applied either backwards or forwards in time.
Zel'dovich approximation \citep{Zel'dovich1970} has been classically used to trace
the density field back in time to the linear regime \citep{NusserDekel1992,
  Doumler2013, Sorce2014}. Recently, several studies
\citep{JascheWandelt2013, Kitaura2013,  WangMoYang2013,
  HessKitauraGottloeber2013, WangMoYang2014} have implemented forward
dynamical models within Markov Chain Monte Carlo algorithms. These
bayesian methods gradually adjust the phases and amplitudes of the ICs
until the evolved density field closely matches the one
traced by the observed galaxy population. We refer to \citet{WangMoYang2014, WangMoYang2016} for a more detailed discussion on the advantages and shortcomings of the various methods.

\setcounter{footnote}{0}

The ELUCID project (Exploring the Local Universe with the reConstructed Initial Density
field) consists in using the large amount of data gathered by the SDSS
(Sloan Digital Sky Survey) to map the local universe and generate
reliable simulations reproducing the distribution of large scale
structures in that survey.
The first large ELUCID simulation\footnote{Additional information about the current ELUCID run can
  be found at \url{http://gax.shao.ac.cn/ELUCID/}.}  was completed in 2015:  it
corresponds to a WMAP 5$^{th}$ year cosmology with $\Omega_M=0.258$,
$\Omega_\Lambda=0.742$ and $h=0.72$.
The simulation volume is a box of  500 Mpc/h per side, where matter is
described by $3072^3$ particles corresponding to a  mass resolution
of  $3.1\times10^8$ \msunh. A more detailed
  background of the ELUCID project is given in Appendix \ref{sec:
    ELUCID}.

One of the main applications of the ELUCID simulation is to build
  realistic mock catalogs by applying various SAMs
to the reconstruction. The validity
of the model can be then tested by direct comparison with the original
observed SDSS sample and group catalog derived using the method of
\citet{Yang2005a, Yang2007}. The first problem is to build a framework for a direct
comparison. This means devising a reliable criteria for identifying
the most likely reconstruction of a given observed group. The second
problem is to differentiate between the effects of the reconstruction
method and those of the SAM.

In \citet{WangMoYang2014}, a forward method that employs the
Hamiltonian Markov Chain Monte Carlo (HMC) algorithm and the PM
dynamics was developed.
This method was tested with $N$-body simulations in order to find the
best set of parameters.  The main criteria was the similarity of the
respective
initial power spectra and of the halo mass functions of an original Dark Matter only
simulation and its reconstruction. The simulation sample they used is independent of any
baryonic physics and observational bias.  For this reason, this
simulation data set may be used to prepare the analysis of the ELUCID
simulations by exploring the limitations of the reconstruction
technique itself.

In this paper, we make use of part of the $N$-body simulations sample from
\citet{WangMoYang2014}. More specifically we use halo catalogs
extracted from these simulations. We first devise a method to match
halos in the original simulation to halos in the reconstructed
one. This catalog of matched halo pairs can be used to test
alternative criteria that can be applied to the ELUCID
simulations . Additionally various internal properties of halos, such as shape,
orientation or angular momentum can be compared. More precisely, given
a reconstructed halo, we would be able to identify which of its
properties were most likely to be constrained by the reconstruction.
Furthermore, $N$-body simulations contain the full formation history of
the halos they describe. By applying this cross-identification of halos
in the original and reconstructed simulations at different times, one
can assess how much of the formation history has been reproduced by
the reconstruction method. Since both internal properties and
formation histories of halos can have a strong impact on the SAM,
these questions are extremely relevant to the ELUCID project.

We detail the method used in the paper in section \ref{sec:method},
before focusing on the results in section
\ref{sec:results}. We discuss our results after a short summary in section \ref{sec:discuss}.

%==============================================================

\section{Method}
\label{sec:method}

In \pararef{sec:sim}, we introduce the $N$-body
  simulation sample and explain how it was processed to obtain the halo catalogs. We describe in detail in \pararef{sec:alcomp}, the halo
  cross-comparison method.

\subsection{Simulations}
\label{sec:sim}

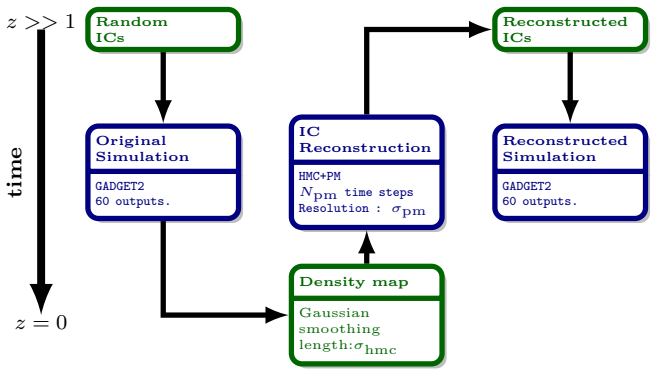
\begin{figure}[ht]
  \begin{tikzpicture}
    \path (-1.5\hu,0) node (IC0) [datas]{{\bf Random \\ICs}
    };

    \path (-1.5\hu,-2\vu) node (Sim0) [codes]{{\bf Original\\Simulation}
      \nodepart{second}
      \tt{GADGET2}\\
      60 outputs.
    };

    \path (0,-2\vu) node (PM) [codes]{{\bf IC\\Reconstruction}
      \nodepart{second}
      \tt{HMC+PM}\\
      $N _{\rm pm}$ time steps\\
      Resolution : $\sigma_{\rm pm}$
    };

   \path (1.5\hu,0) node (ICr) [datas]{{\bf Reconstructed\\ICs}
    };

    \path (1.5\hu,-2\vu) node (Simr) [codes]{{\bf Reconstructed\\Simulation}
      \nodepart{second}
      \tt{GADGET2}\\
      60 outputs.
    };

    \path (0,-4\vu) node(smoothd) [datas,rectangle split, rectangle
    split parts=2]{{\bf Density map}
      \nodepart{second}
      Gaussian smoothing \\
      length:$\sigma_{\rm hmc}$
    };
    \draw[->,line width=2pt,draw=black] (IC0.south) -- (Sim0.north);
    \draw[->,line width=2pt,draw=black] (Sim0.south) |- (smoothd.west);
    \draw[->,line width=2pt,draw=black] (smoothd.north) -- (PM.south);
    \draw[->,line width=2pt,draw=black] (PM.north) |- (ICr.west);
    \draw[->,line width=2pt,draw=black] (ICr.south) -- (Simr.north);
    \draw[->,line width=3pt,draw=black]  (-2.4\hu,0) -- (-2.4\hu,-4\vu);
    \path (-2.6\hu,-2\vu) node[rotate=90] {\bf time};
    \path (-2.4\hu,-4.1\vu) node {$z=0$};
    \path (-2.4\hu,0.1\vu) node {$z>>1$};

\end{tikzpicture}
\caption{This diagram introduces how a pair of original/reconstructed
  simulations were generated. We start from initial conductions
  randomly generated from the initial power
  spectrum. The original
  $N$-body simulation is produced from these initial conditions down to redshift 0. The last output is
  processed as to produce a new set of initial conditions. This new set is used to run
  the reconstructed $N$-body simulation.
  \label{fig:intro_constrain}}
\end{figure}

We summarize in \figref{fig:intro_constrain}  the processes involved in
creating a simulation pair. The first one (referred as the
  original) is built from random initial conditions
(generated from the power spectrum predicted for the cosmology), the second
simulation is a reconstruction of the first.
The initial conditions of the reconstructed simulation
are generated following \citet{WangMoYang2014}.  As the schematic
shows, we proceeded as follows:

\begin{enumerate}
\item The original simulation is run from random initial conditions.
\item The final (redshift 0) output is post-processed with a gaussian smoothing
  kernel to produce the density field.
\item This density map is used as input for the reconstruction
  method to produce a reconstruction of the initial conditions.
\item The reconstructed simulation is run from the reconstructed initial conditions.
\end{enumerate}

We list in \tabref{tab:sims}, the simulation data set used in
this paper. Each simulation has $512^3$ particles,
of mass $2\times 10^{10}$ \msun, in a comoving  box of side  300 \mpch.
The cosmological parameters are $\Omega_M=0.258$,
$\Omega_\Lambda=0.742$ and $h=0.72$. The cosmology is not relevant, as
we focus rather on numerical techniques than on the
effect of the cosmological model. We emphasize, that all simulations are run with the same $N$-body
code, here {\tt GADGET2} \citep{Springel2005}, with the same
time outputs (60 outputs equally spaced in $\Delta\log(1/1+z)$
from $z=18$ to $z=0$). The only difference between simulations mentioned in this
paper is in the initial conditions.

\begin{table}[ht]
  \centering
  \begin{tabular}{l l l l}
    Name&$\sigma_{\rm HMC}$ [Mpc/h]&$\sigma_{\rm PM}$ [Mpc/h]&N$_{\rm PM}$\\%&ID\\
    \hline
    L300A&&&\\%&L300\\
    \hline
    L300Ac1&2.25&1.5&10\\%&L300Nd20002\\
    L300Ac2&3   &1.5&10\\%&L300Nd20012\\
    L300Ac3&4.5&1.5&10\\%&L300Nd2002\\
    \hline
    L300B&&&\\%&L3002\\
    \hline
    L300Bc1&2.25&1.5&10\\%&L3002Nd20002\\
    L300Bc2&3   &1.5&10\\%&L3002Nd20012\\
    L300Bc3&4.5 &1.5&10\\%&L3002Nd2002\\
    \hline
  \end{tabular}
  \caption{List of $N$-body simulations. We have two series
      of simulations L300A and L300B. The original simulations are
      L300A and L300B. From each of the original simulations, 3
      reconstructions where performed. The extensions c1, c2 and
      c3 is used to distinguish them from their respective
      original and refer to the corresponding sets of reconstruction
      parameters ($\sigma_{\rm HMC}$, $\sigma_{\rm PM}$, N$_{\rm PM}$)  indicated in the row. \label{tab:sims}}
\end{table}

As indicated in \tabref{tab:sims}, we have 2 independent
  $N$-body simulations: L300A and L300B. For each of these original
  simulations 3 sets of reconstructed simulations were generated. The
  reconstruction method is detailed in \citet{WangMoYang2014}. It
  depends on the density smoothing
  $\sigma_{\rm HMC}$ and the Particle Mesh (PM) parameters: N$_{\rm PM}$
  (number of steps) and $\sigma_{\rm PM}$ (grid length). The smoothing scale had to be chosen to reduce the
    resolution effects of the PM model. Choosing  $\sigma_{\rm HMC} \leq
    \sigma_{\rm PM}$ leads to significant discrepancies in the 2 point
    correlation functions at small scales \citep[see
    Figure 3][]{WangMoYang2014}. The smallest  $\sigma_{\rm HMC}$ we
   consider is  $1.5\times\sigma_{\rm PM}$, while
   $2\times\sigma_{\rm PM}$ renders these discrepancies
   negligible. The choice of $3\times\sigma_{\rm PM}$ represents a compromise between further reducing this resolution effect and limiting the loss of information at small scales.
In this paper we use the simulation sample that is used to explore the
  impact of the density smoothing with fixed PM
  parameters. We rather
  focus halo catalogs extracted from these simulations than the raw
  simulations data.

There are a large variety of group-finding
algorithms, starting from simple halo finding methods such as the friend of friend
\citep[FOF]{Davis1985} or Spherical Over-Density
\citep[SOD]{LaceyCole1994}, through more elaborate subhalo
  finders such as {\tt SUBFIND}
\citep{Springel2001b},  {\tt AdaptaHOP} \citep{Aubert2004,Tweed2009}, or
 {\tt AHF} \citep{Knollmann2009} to even more complex ones such as phase-space group-finders
{\tt 6DFOF} \citep{Diemand2006}, {\tt HST} \citep{Maciejewski2009},
or {\tt RockStar} \citep{Behroozi2013} \citep[see
review by][and
subsequent ``subhalos going notts'' papers]{Knebe2011,
  Onions2012}. Aside from the algorithm
used to collect particles, the properties associated with the halos
are not necessarily standardized. The position can be defined
as the center of mass of the collection of particles,  the
position of the density peak or the position of the most
bound particle. As mentioned by \citet{Cui2016},   the two positions are normally aligned with each other.
The
mass itself can be defined as the total mass of the collection of particles, or as the virial
mass defined as the total mass within a spherical (or ellipsoidal) region. This
  virial region can be defined as an over-density of either 200$\times
\overline{\rho_{\rm DM}}$,   200$\times \overline{\rho_{\rm crit}}$ or $\Delta(z) \times
\overline{\rho_{\rm crit}}$. These criteria also lead to multiple definitions of halo size,  boundary and shape.

For simplicity, we have used the standard
friend of friend \citep[FOF][]{Davis1985} algorithm where groups
  are built by iteratively linking particles to all its neighboring
  particles closer than  a
specific distance. In our case, the distance used is $0.2 \times \overline\Delta_N$ ($b=0.2$)
where $\Delta_N=L_{\rm box}/{N^{1/3}}$ is the mean inter-particular
distance of the $N$-body simulation in a volume of side $L_{\rm box}$
($N^{-1/3}$ in box units).  We also mention that no unbinding procedure have been
applied to the halo catalogs.  This group-finder is often the baseline for
more complex ones.  If we were not to find any match between FOF halo catalogs,
it is unlikely we would find any between catalogs build using other
group-finders. Internal properties comparison and subhalo
cross-identification can still be implemented at a later stage for FOF cross-identified halos.

Our starting point here for each simulation is a list of
  particles with their associated indexes, positions, velocities and mass. We
  also have a list of halos defined as a collection of
  particle indexes. From this collection we can easily calculate the
  position, mass or other useful quantities relevant to the halo.

\subsection{Halo cross-identification}
\label{sec:alcomp}

Having introduced the data we use in this paper, we
  now describe the comparison algorithm.

\subsubsection{General idea}

\figref{fig:sym_comp} illustrates the initial layout of
  a halo cross identification method applied to two catalogs A and B.
\begin{enumerate}
\item We compare halo catalog B to halo catalog A:
  \begin{itemize}
  \item Each halo in catalog B can be associated to at most one halo in catalog
  A. Multiple halos B$_1$, B$_2$... can be associated to
    the same halo A$_i$ in catalog A (upper half of \figref{fig:comp_step1}).
  \item Only one of
      these halos (B$_1$, B$_2$...) can be ``cross-identified'' with
      A$_i$, the others remain ``associated'' with A$_i$ (upper half of \figref{fig:comp_step2}).
  \end{itemize}
\item We compare halo catalog A to halo catalog B, following
    the same method to ensure consistency (lower halves of Figures
    \ref{fig:comp_step1} and \ref{fig:comp_step2}).
\end{enumerate}

There are multiple reasons for us to design the method so that
  the comparison is run in both directions. The first reason is not to
  introduce any bias to our interpretation by implementing a
  fiducial. Let's suppose that for each halo A, the best match found in B is
systematically larger. We cannot rule out that this systematic is not
caused by the matching criteria unless we reverse the problem.  Running
the comparison both ways, provides the means to ensure that the
cross-identification is consistent.
The second reason is the interpretation of the non cross-identified
halos. To illustrate this, we now suppose that catalog A is a subset of
catalog B. By comparing catalog B to catalog A, we would find that every
halo in B is cross-identified. We would then conclude that catalog A and B
are identical.  But by comparing catalog A to catalog B, we would find a
population of halos in catalog B that are not cross-identified.  We
would draw the correct conclusion; catalog A and B
are different.

\begin{figure}[ht]
  \centering
  \subfloat[We search for possible matches (single-tip
    arrows) of halos B in
    catalog A (top) and for possible matches of halos A in
    catalog B (bottom). This constitutes a mapping of all associations.\label{fig:comp_step1}]{%
    \includegraphics[width=\onecolfigwidth]{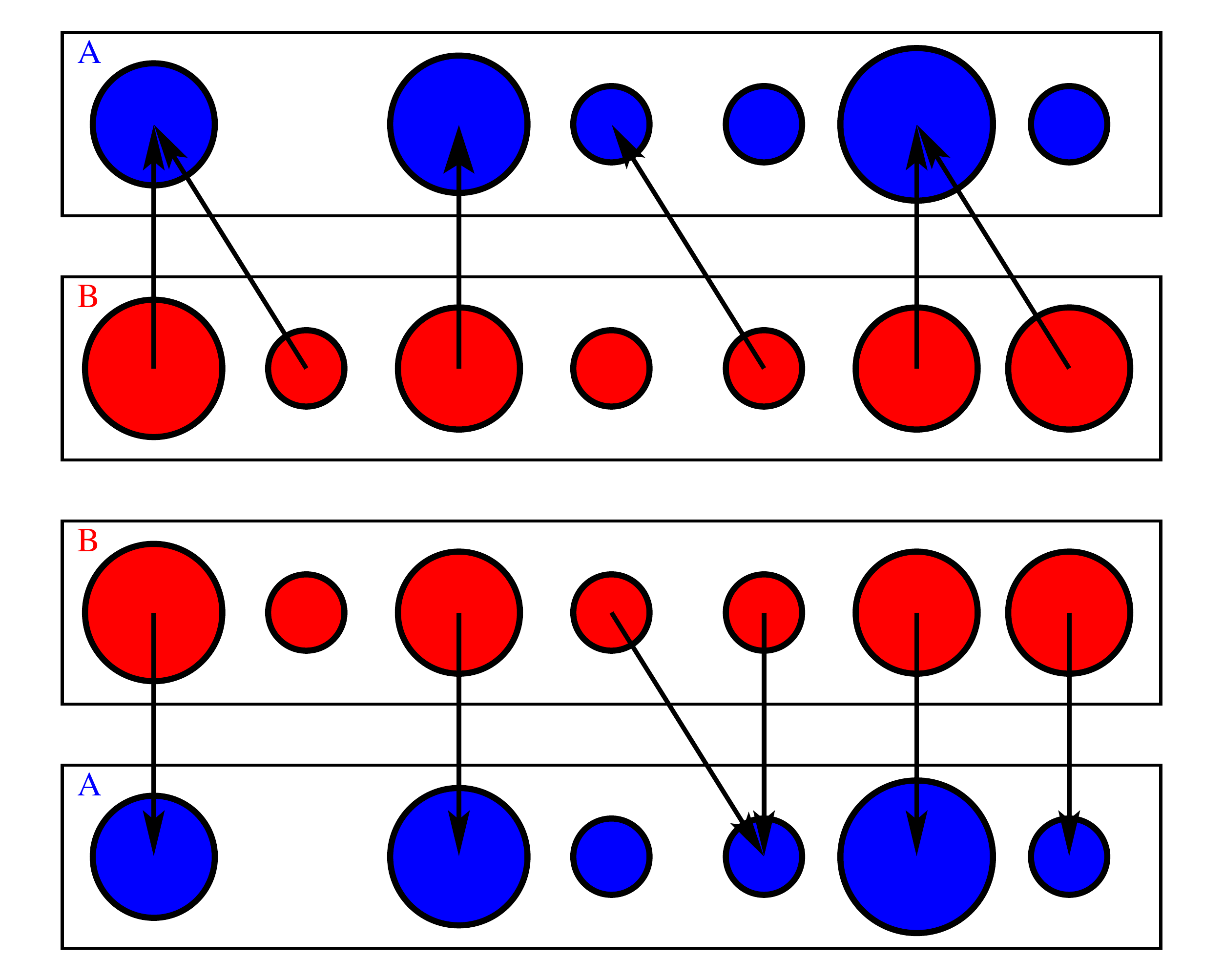}%/comp_fig3.pdf}
  }\\
  \subfloat[Top: Among possible matches of halos A in catalog B one
    is selected as cross-identification (double-tip arrows), the
    remaining associations are still indicated. Bottom: the same selections is applied for matches of
    halos B in catalog A.
  \label{fig:comp_step2}]{%
    \includegraphics[width=\onecolfigwidth]{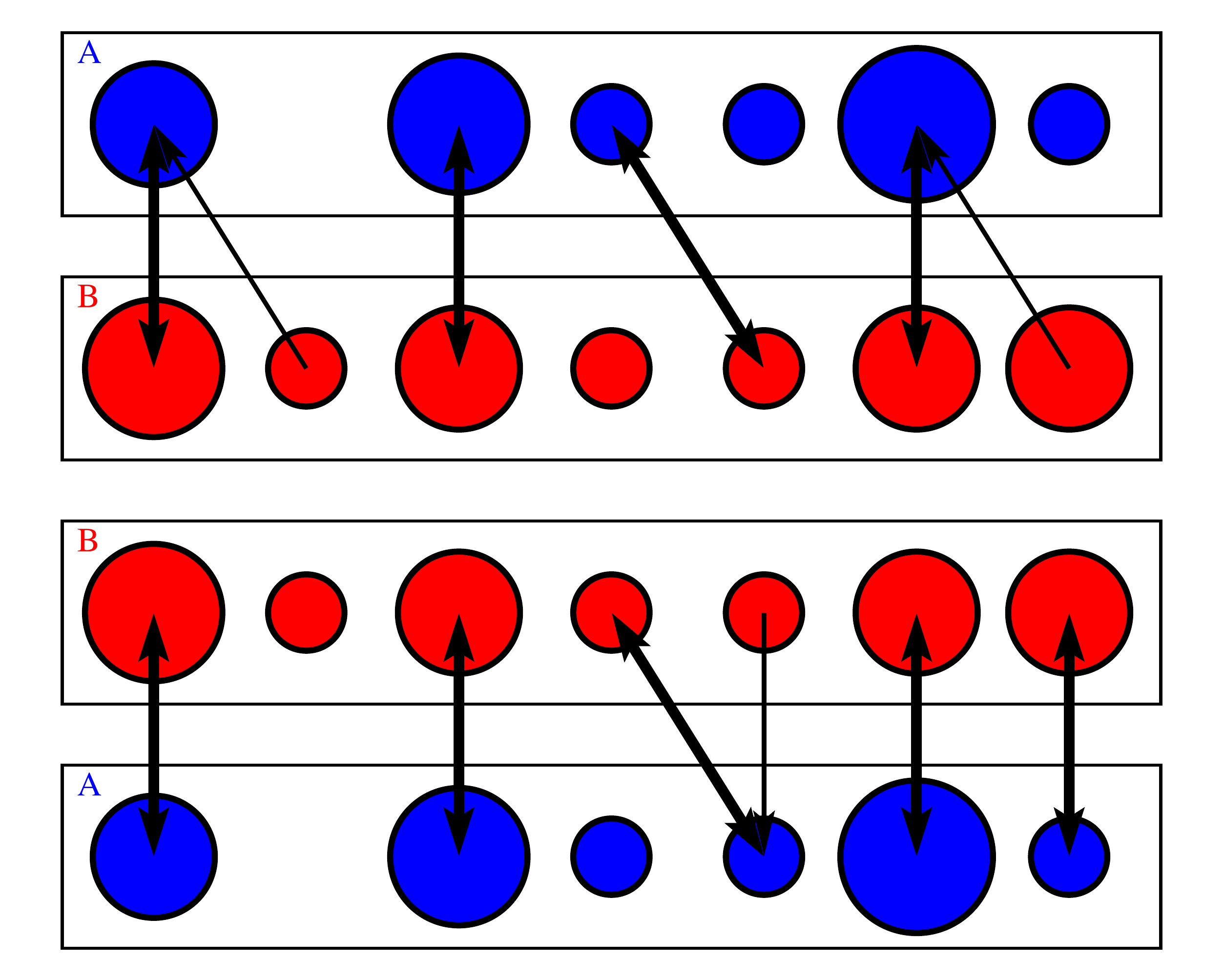}%/comp_fig4.pdf}
  }
  \caption{Illustration of the cross-identification method, the
      method is applied twice from catalog B to catalog A and from catalog A to catalog B.
    \label{fig:sym_comp}}
\end{figure}

As detailed in \figref{fig:sym_comp}, a one way identification can
provide multiple correspondences for a halo. If we only care about
cross-identification (double tip arrow), we could choose one
possibility and discard the others. We chose however to keep track of these discarded
choices as associated halos (single tip arrow) and distinguish them from halos that have
no possible counterparts. In order to collect these associated halo
populations from both catalogs, it is also necessary to run the comparison both
ways.

\begin{figure}[ht]
  \centering
  \includegraphics[width=\onecolfigwidth]{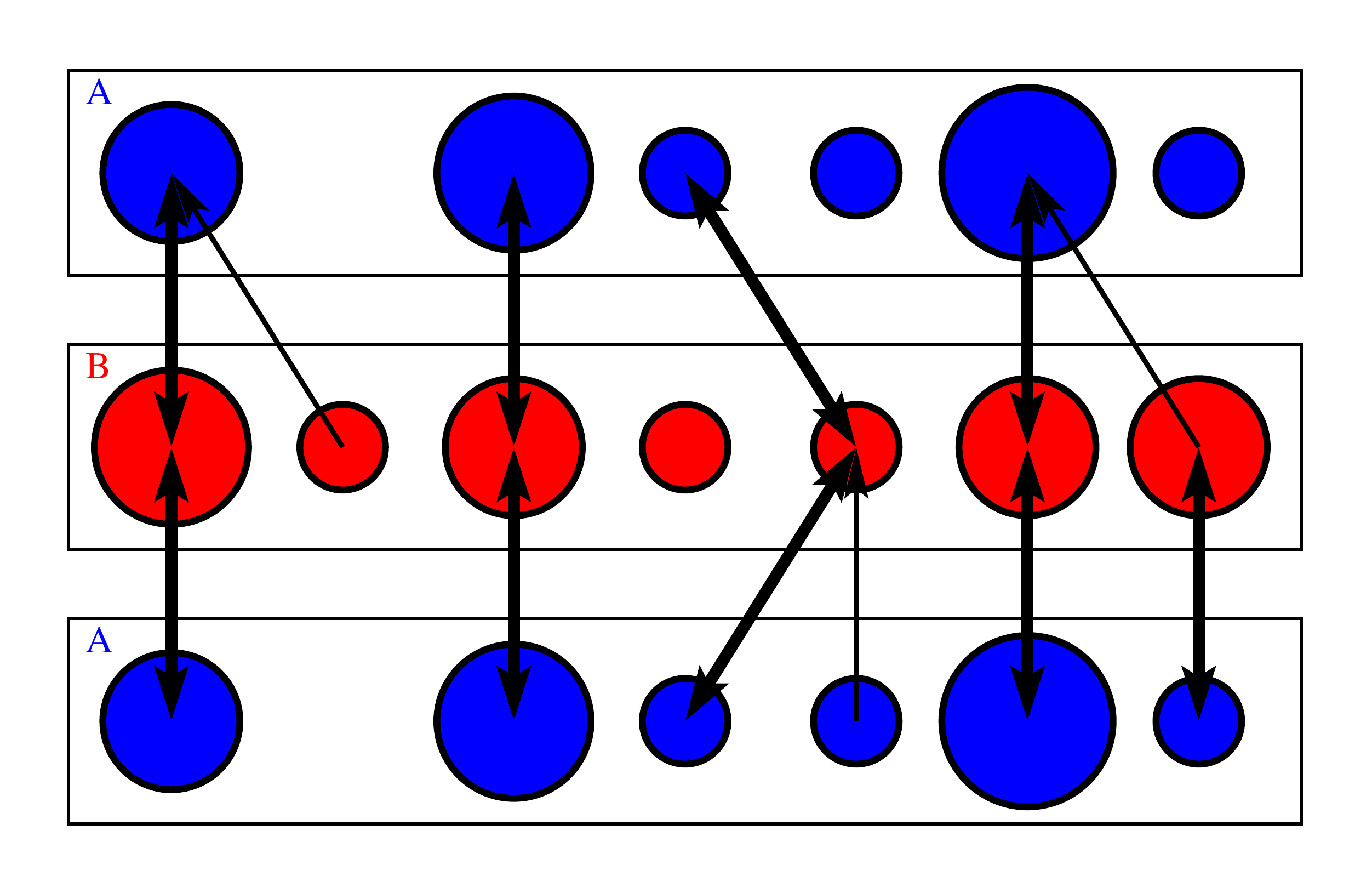}%/comp_fig7.pdf}
  \caption{Code optimization: The layout of the algorithm is
      reorganized,   lower
      part of inverted \label{fig:comp_step4} and combined to the top
      one. The same cross-identification search can be implemented in
      both directions as one run.
    \label{fig:tree_step}}
\end{figure}

As we have described it, our algorithm need to be run twice,
once from B to A and once from A to B. The results have then
to be compared for possible mismatches.
One very simple improvement, is that instead of reading/using both catalogs
twice, the code can be restructured to read/use only one of the
catalogs twice and the other once. The procedure described in
  \figref{fig:sym_comp} becomes that illustrated in \figref{fig:tree_step}. We note here that we
used a A-B-A configuration, but a B-A-B configuration would provide the same answers. The structure of the
  cross-comparison between catalogs takes the form of a tree, with one
  main branch (trunk) for cross-identification and at secondary
  branches for associations.

\subsubsection{Cross-identification criteria}

Since the aim of
reconstruction is to obtain the same spatial
distribution of large scale structures, the most logical choice would
be to compare the position of halos. Minimizing the distance may mislead the
  algorithm into pairing up halos of very different masses,
  while the correct reconstructed halo could be found at a slightly
  larger distance.  If one wanted to focus on the most massive structures only, one could
simply, cross-identify halos starting from the high mass end of the mass function. But this cheap and easy method would
quickly break down after a few 100 groups.
Solving the problem requires to find a criteria that would
  minimize both differences in positions and mass between original and
reconstructed halos matches. \citet{WangMoYang2016} implemented
such a method to associate galaxy groups to halos in the ELUCID
reconstructed simulation.  A future cross-comparison method between real
and reconstructed data will implement similar criteria within the
framework we propose in this paper.  The purpose of this paper is to
build reliable cross-identified catalogs in order to evaluate and
calibrate such a method.

The reason why we use reconstructions of $N$-body simulations
  is that halos properties are no longer restricted to position,
  mass and size. Furthermore halo cross-identification is a known
  problem for simulations. Building a merger-tree consists in
  associating halos between time outputs. Testing a new group-finder
  also requires to be cross-compared with a halo catalog built using an
  other algorithm. In both case, cross-identification is applied for
  one simulation (e.g.  one set of  initial conditions) and particles indexes
  are used as tracers. The problem is solved by
  optimizing the quantity of particles in common matched halos.
But having different initial conditions implies that identical particle indexes
have different displacement fields applied to them. Technically,
particle indexing has no effect on the simulation outcome, and should
be entirely random.

However, in $N$-body simulations, particle
indexing is not random. Indexes are assigned following a fixed grid
to which the displacement field characterizing the initial conditions is applied. Thinking back to the problem
at hand, one may suppose that if the reconstruction proves to be
truly efficient, reconstructed halos would appear out of similar
regions in the initial conditions.
Provided that the same particles indexes are associated to the
same coordinates in the initial conditions, we could expect that a perfectly
reconstructed halo would have the same particle indexes as the
original one.  In the case of $N$-body simulations, this means we can
apply a tree-builder to the problem.

We can
  illustrate this by supposing that catalog B is the earlier
  output. For each halo in catalog B we first map out a
  list of all possible descendants in catalog A. Each candidate in
  A has at least one particle in common with halo B, so their common
  mass is $M_{A \cap B} > 0$. This constitutes a  {\it graph}.
To create a {\it merger-tree} out of the {\it graph}, one need to
select one {\bf single} descendant, and update the list of
their progenitors accordingly.
In tree-builders, the strength of connections between progenitors and
  descendent is weighted by a {\it merit function}. The most widely used is the normalized shared merit
function, and for two halos of masses $M_A$ and $M_B$ is defined
as:
\begin{equation}
  \mathcal{M}_m(A,B) =\frac{M_{A \cap B}^2}{M_A\times M_B} \,.
  \label{eq:merit}
\end{equation}
For each halo in catalog B, the descendant
  in catalog A with the highest merit is selected.  Among the
  progenitors found in catalog B, the main progenitor of any halo in
  catalog A is the one with the highest merit. Going back to the cross-identification problem, {\it cross-identified} would be
translated as  {\it main progenitor} and  {\it associated} as
{\it secondary progenitor}.

As the review by
\citet{Srisawat2013} shows, many tree-builders are available
and most are variations of the algorithm by
\citet{LaceyCole1994}. They differ in the way some
very technical corrections are implemented. \citet{Avila2014}
demonstrate that the choice of group-finder will have a stronger impact on the merger-tree themselves than the
choice of tree-builder.  The tree-builder
used here is {\tt TreeMaker} \citep{Hatton2003, Tweed2009}.  Using this
tree-builder simply requires that the original (as A) and
reconstructed (as B)
halo catalogs are used as indicated in \figref{fig:tree_step}.

This tree-structure is useful for
validating the cross-identification. The main branch should
  start and end with the same halo and secondary branches should
  contain at most one halo.
As one can see from our illustration \figref{fig:comp_step4} (or even
\figref{fig:comp_step2}), this is not necessarily the case. These inconsistencies could represent a problem. In Appendix
  \ref{sec:cor_comp}, we describe how these inconsistencies are solved before constructing
the cross-identified and associated catalogs.
Some additional selection is applied for these catalogs.  By default we apply a merit threshold of 0.25, but
we may increase or decrease this parameter to derive more or less complete
 or strictly accurate cross-identified catalogs.  We chose this value since it can
 be interpreted as a minimum quadratic average of 50\% common mass
 between two cross-identified groups.
For improving the
 interpretability of the associated catalogs, we systematically use,
 after cross-identification, a
 threshold of 50\% of the common mass ($M_{A \cap B}/M_A > 0.5$ for the associated
 catalog A, $M_{A \cap B}/M_B > 0.5$ for the associated
 catalog B). Even though their merit is below the first threshold,
 discarded cross-identified halos can be added to the
 associated catalog if their common mass above that second threshold.

%==============================================================

\section{Results}
\label{sec:results}

We assess in \pararef{sec:smooth} the effect
of the density smoothing scale on the reconstruction. We further test
in \pararef{sec:testth} the compatibility of the mass based
  merit we use to a distance based merit function. The impact of the
the merit threshold selection on the cross-identified  sample
is tested in \pararef{sec:Mthreshmf}. The evolution
  of the quality of the reconstruction at earlier redshift is explored
in \pararef{sec:compz}. We check in \pararef{sec:compzdz}, whether
earlier time steps of the reconstructed simulations could correspond to more
accurate reconstructions.

\subsection{Density smoothing and high mass bias}
\label{sec:smooth}

\begin{figure*}
  \centering
  \subfloat[$\sigma_{\rm HMC}=2.25$\label{fig:mf_300Ac1}]{%
    \includegraphics[width=\thirdtwocolfigwidth]{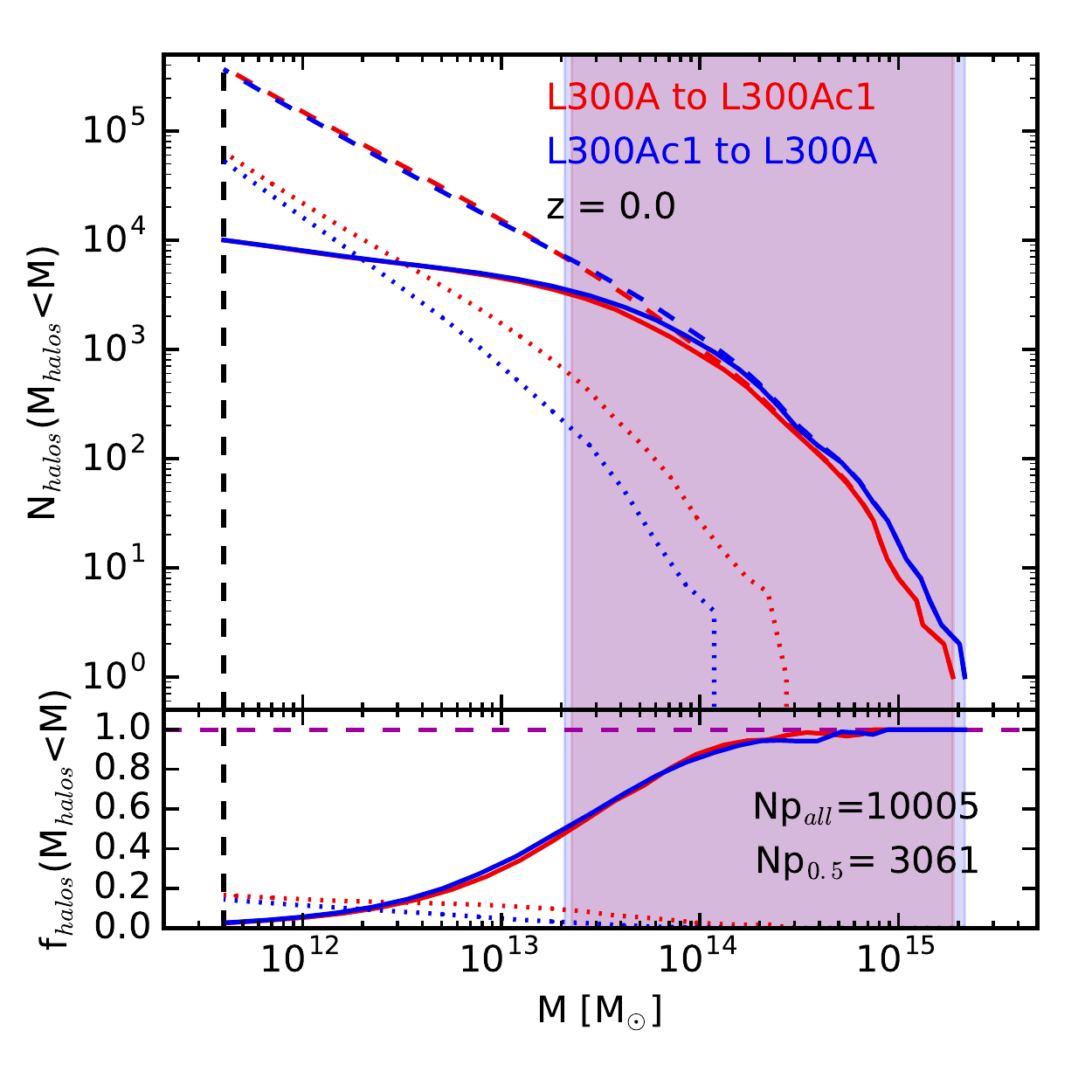}%/cummf_L300A_vs_L300Ac1_060.pdf}
  }
  \subfloat[$\sigma_{\rm HMC}=3$\label{fig:mf_300Ac2}]{%
    \includegraphics[width=\thirdtwocolfigwidth]{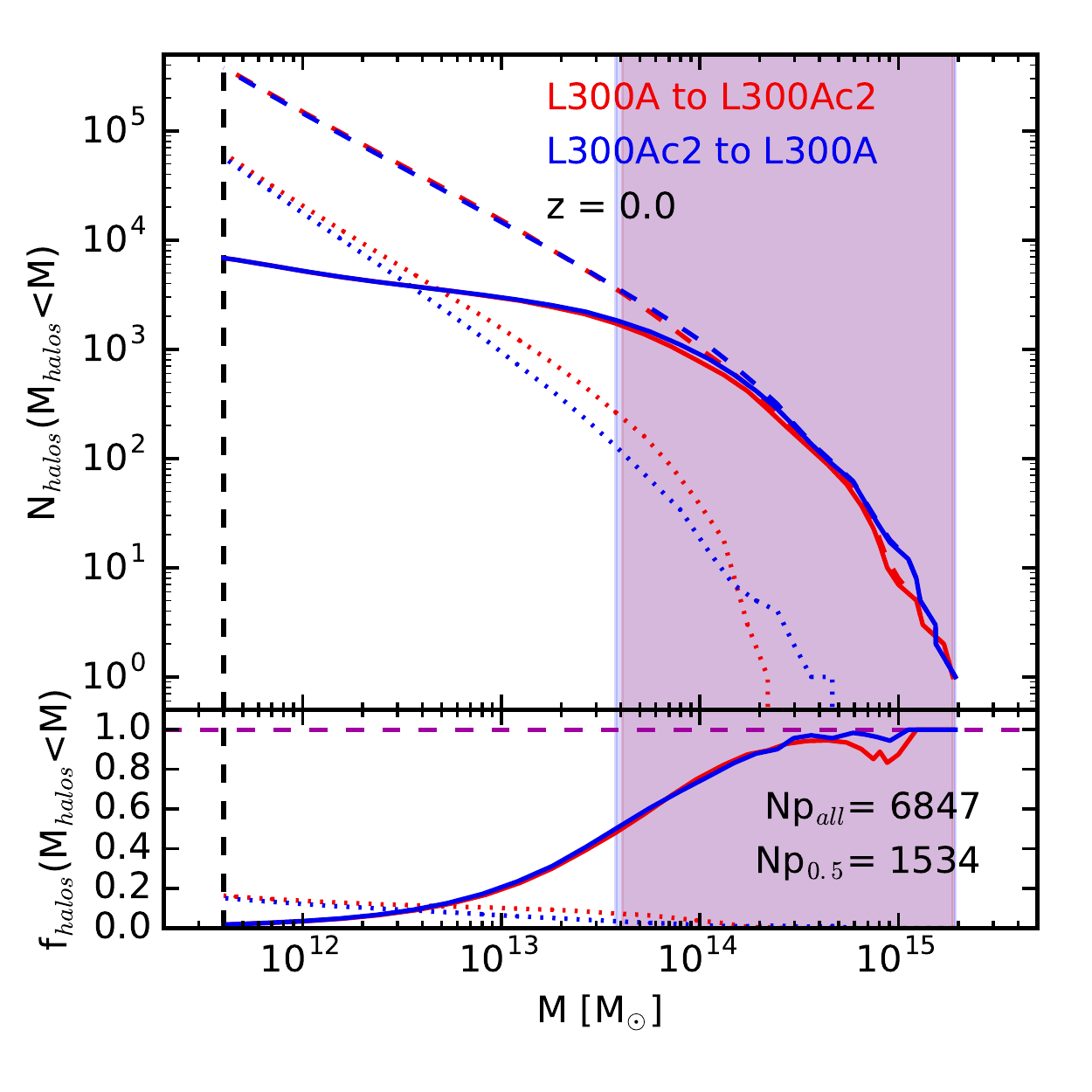}%/cummf_L300A_vs_L300Ac2_060.pdf}
  }
  \subfloat[$\sigma_{\rm HMC}=4.5$\label{fig:mf_300Ac3}]{%
    \includegraphics[width=\thirdtwocolfigwidth]{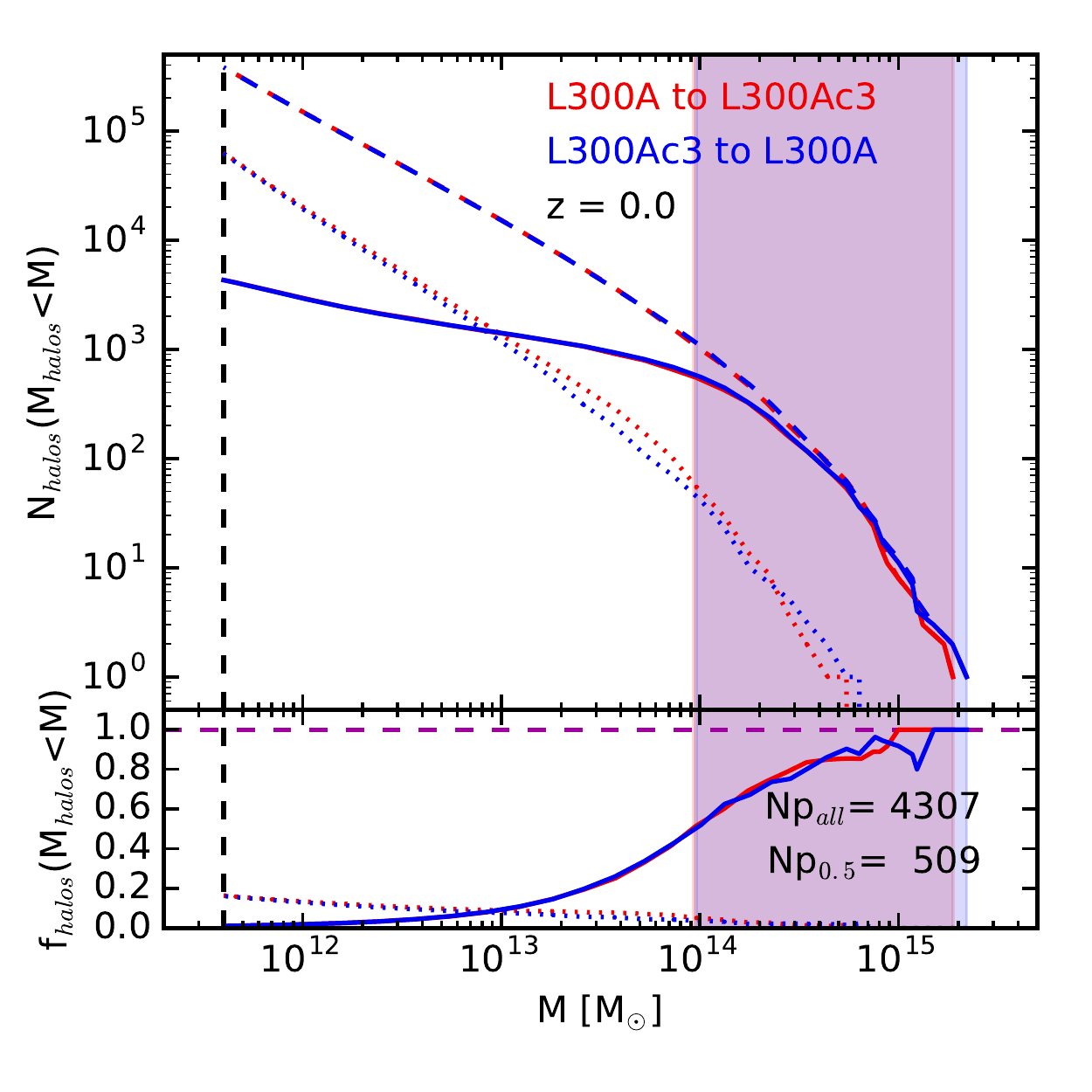}%/cummf_L300A_vs_L300Ac3_060.pdf}
  }
  \caption{Each subfigure assesses the performances
       of a different reconstruction of the same original simulation. {\it Top panels:} For both original (in red) and
    reconstruction (in blue) we show the cumulative mass functions of all
    halos (dashed curves), of the cross-identified subsamples (solid curves) and the
    associated samples (dotted curves). {\it Bottom panels} The last four curves are shown
    again after normalization by the cumulative mass functions
    indicated by
    the  horizontal purple dashed line. Np$_{\rm all}$ denotes the
    total number of cross-identified pairs, and Np$_{\rm 0.5}$ the
    number of pairs for which the reconstruction is 50\% complete. In
    both panels, additional shaded regions indicate the 50\% completeness
    regions, and the vertical dashed line the minimal halo mass. \label{fig:mf_smooth}}
\end{figure*}

\begin{figure*}
  \centering
  \subfloat[$\sigma_{\rm HMC}=2.25$\label{fig:md_300Ac1}]{%
    \includegraphics[width=\thirdtwocolfigwidth]{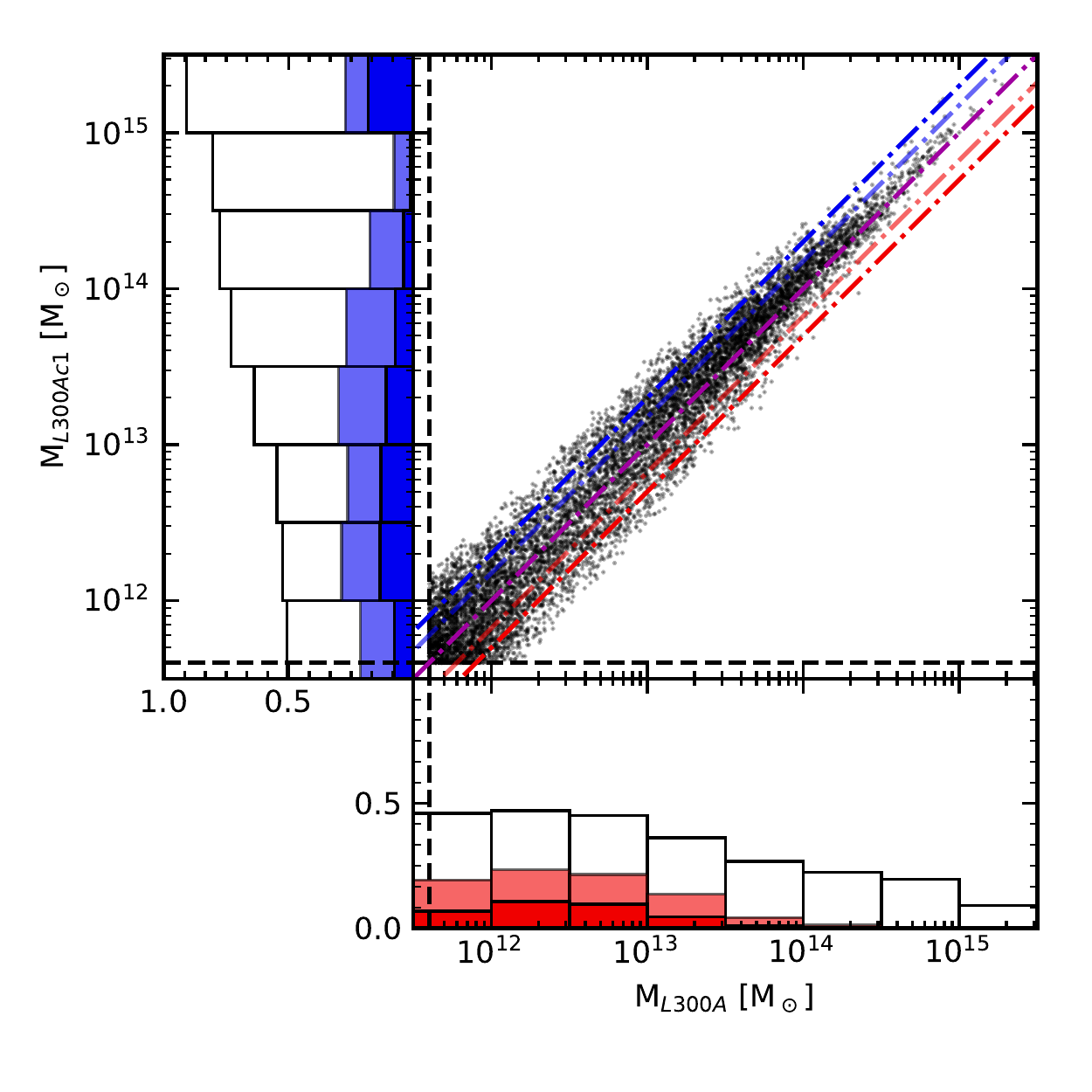}%/mass_test_L300A_vs_L300Ac1_060.pdf}
  }
  \subfloat[$\sigma_{\rm HMC}=3$\label{fig:md_300Ac2}]{%
    \includegraphics[width=\thirdtwocolfigwidth]{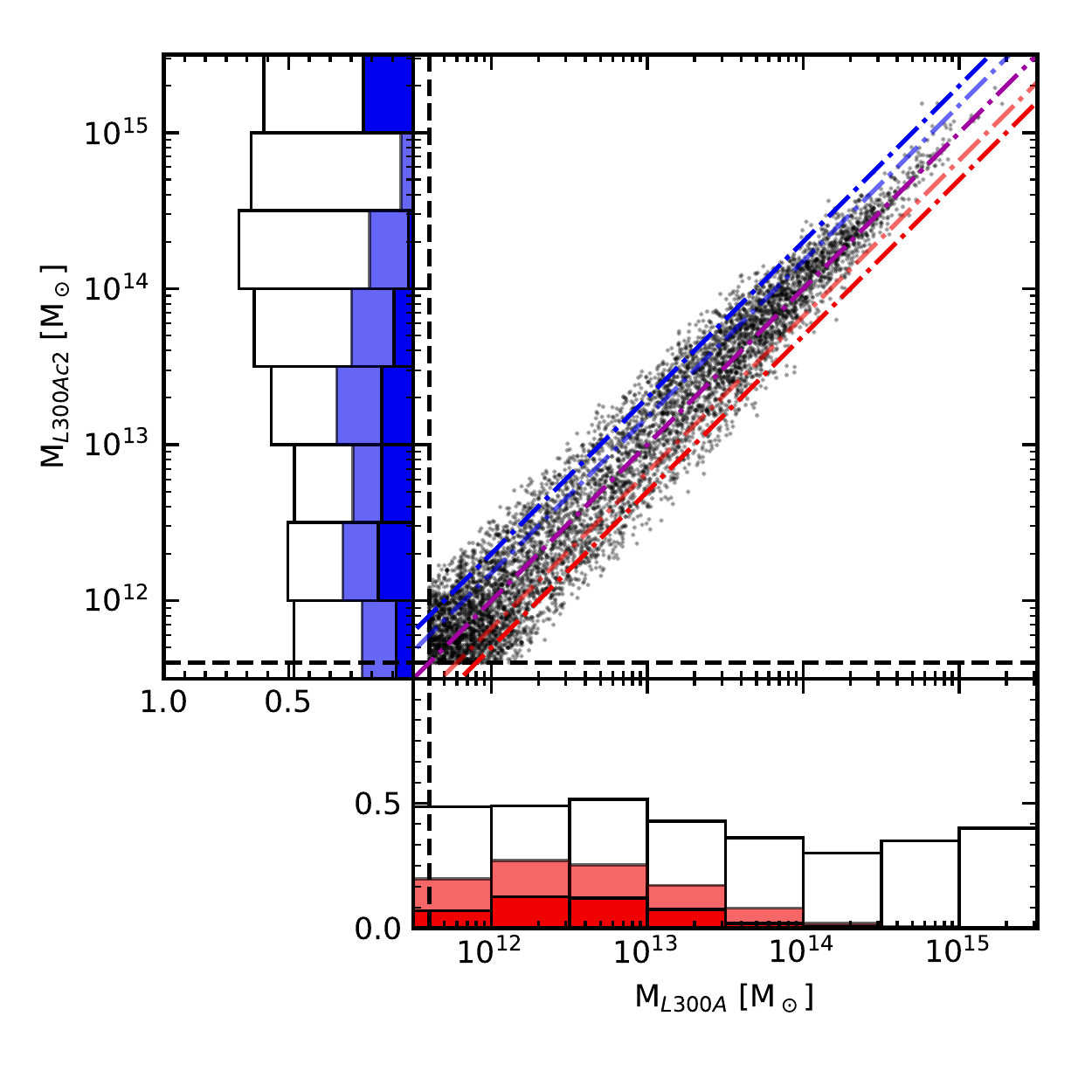}%/mass_test_L300A_vs_L300Ac2_060.pdf}
  }
  \subfloat[$\sigma_{\rm HMC}=4.5$\label{fig:md_300Ac3}]{%
    \includegraphics[width=\thirdtwocolfigwidth]{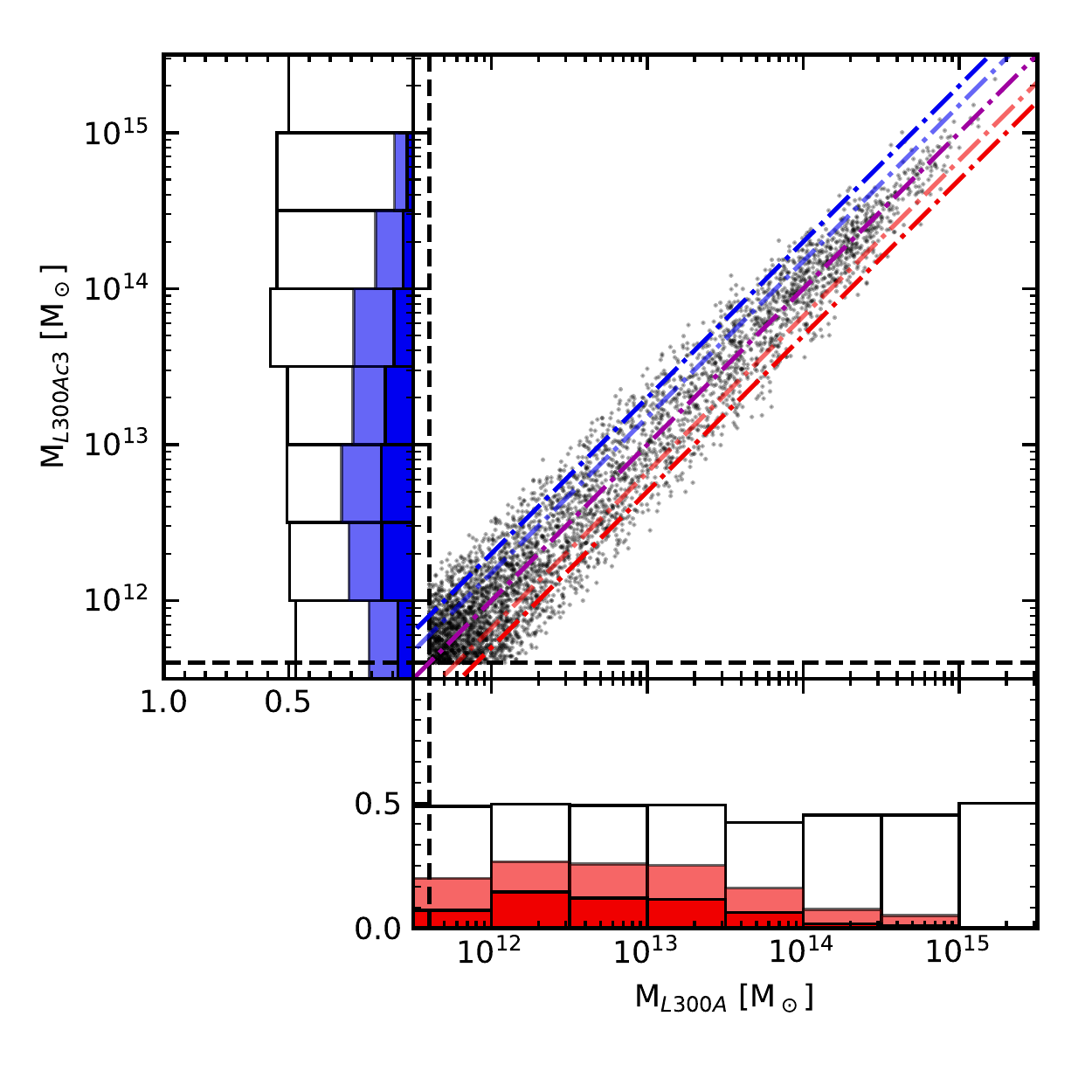}%/mass_test_L300A_vs_L300Ac3_060.pdf}
  }
  \caption{Each subfigure assesses the performances
      of a different reconstruction of the same original simulation.
      The scatter plot represents for each
      cross-identified halo pair, the mass in the reconstructed
      simulation (ordinate) as a function of the mass in the
      original simulation. The 1:1 relation is marked as a purple dot-dashed
      diagonal line. The fraction of data points above and below this line are represented
      as a function of the medium mass  as black histograms
      respectively on the left panel (over-estimation in the
      reconstruction) and the bottom panel (over-estimation in the
     original) respectively. Additional dot-dashed diagonals
     represents  different mass ratios between reconstruction and original, from
      top left to bottom right: 2 (blue), 1.5 (light blue), 1/1.5
      (light red)  and 1/2 (red). The fractions above the blue lines
      (large overestimations in the reconstruction) and below the red
      ones (large overestimations in the original)
are also represented in the horizontal (left) and vertical (bottom)
      histograms respectively following the same color code. In all panels vertical
      and horizontals black dashed lines represent the halo mass resolution.
     \label{fig:md_smooth}}
\end{figure*}

\figref{fig:mf_smooth} illustrates the effect of the HMC smoothing
  length on the reconstructed mass function at the final
redshift. Each subfigure corresponds to a different value of
$\sigma_{\rm HMC}$ used for the reconstruction of the same original simulation.
In the top panels, the cumulative mass functions of each catalog are
represented as dashed curves. In each case, the mass functions of the
reconstruction (in blue) appear quite similar to the original (in
red). Using solid curves following the same color code,  we also show  the cumulative
mass functions of the cross-identified subsample of each simulation.
At the high mass end, these curves are superimposed on the dashed
ones. In figures \ref{fig:mf_300Ac1} and \ref{fig:mf_300Ac2}, where
lower values of $\sigma_{\rm HMC}$ are tested, the reconstructed mass
function is higher than in the original.

This mass bias could be caused by an over-production of massive halos in the
reconstruction or by a tendency towards reconstructing halos with a larger
mass. The former possibility is ruled out since the cross-identified
mass function overlaps the mass function of the complete catalog. The
lower panel further confirms this statement: it displays (as solid
curves) the ratios of the
cross-identified mass functions to the mass functions  of their
respective simulations. We refer to these curves as cross-identified
fractions. A systematic over-production of massive halo in the
reconstruction would translate at the high mass end as a
cross-identified fraction lower than unity
for the reconstruction (blue curve), and a cross-identified fraction
close to unity for the original (red curve).

These panels also illustrate how the reconstruction degrades as we
explore lower masses. Below $10^{14}$ \msunh, the
cross-identified sample diverges from the complete one, effectively
illustrating the increasing lack of completeness of the reconstruction
as we consider lower and lower masses. The cumulative
mass functions of associated halos (associated fractions for
short) is represented as dotted curves (red for the
original, blue for the reconstructions). These associated halo mass functions highlight an additional
  discrepancy between original and reconstructed simulations. For
  low $\sigma_{\rm HMC}$ values, associated mass functions appear to be higher for
the original especially in  the  $[10^{12}, 10^{14}]$ mass
range. Associated halos from the original
simulation that appear in excess may be found within larger halos in the
reconstruction. This trend is likely to be related to the mass excess
in reconstructed halos at the high mass end.

In order to further quantify the quality of the reconstruction, we have
added some extra information to these figures. We indicate the total
number of pairs in the cross-identified sample Np$_{\rm
  all}$. Using both cross-identified fractions, we have estimated  the
mass from which 50 \% of the halo catalog is  reconstructed.
We start by selecting halos pairs corresponding to
  cross-identified fractions greater than 0.5 in both simulations. We
  denote the number of such pairs as Np$_{\rm 0.5}$.
The
respective mass range covered by this 50\% complete cross-identified
sample is represented with shaded regions color coded accordingly to
the simulation with the overlap appearing in purple. As we explore larger smoothing lengths from  Figures
\ref{fig:mf_300Ac1} to \ref{fig:mf_300Ac3}, the total number of
cross-identified pairs Np$_{\rm
  all}$ decreases from $\sim 10^4$ to $\sim 7000$ then $\sim 4300$ for smoothing lengths
2.25, 3 and 4.5 Mpc/h. The corresponding numbers at 50\% completeness
are obviously also lower ($\sim 3000$, $\sim 1500$ and $\sim 500$),
and are related to smaller mass ranges as the minimum
increases ($2\times10^{13}$, $4\times10^{13}$ and $9\times10^{13}$
\msun). Additionally the 50\% completeness mass range found for the
reconstructed simulation appears to be wider in all three
cases. This seems to be related to higher values of the mass function for the reconstruction in the high mass end.

It appears from this figure that
there is a bias towards increasing halo mass during reconstruction at the high mass
end at least. We
  explore this mass increase further  in \figref{fig:md_smooth}. In order to compare directly the
masses of cross-identified halos we
  have produced a scatter plot. Each point corresponds to a
  cross-identified pair, with the
mass as measured in the reconstructed simulation displayed as a function of the
mass as measured in the original one. In order to guide the eye,
various mass ratios are indicated with dash-dotted diagonal lines, with
the 1:1 relation in purple, 50\% and 100\% excess for the reconstructed simulation in
blue, 50\% and 100\% excess for the original one in red. The horizontal histograms
in the left panel show, as a function of the median mass, the
fraction of reconstructed halos with an excess in mass. Excesses
greater than 50\% and 100\% appear in light blue and dark blue. The vertical histograms
in the bottom panel show, as a function of the median mass, the
fraction of original halos with an excess in mass. Excesses
greater than 50\% and 100\% appear in light red and dark red.

Since the mass bias is most prominent for the smallest smoothing length, it
should also appear in \figref{fig:md_300Ac1}.
One can assess qualitatively from the scatter plot that
  there does not seem
to be any mass bias in this first case up to $10^{13}$ \msunh. However
the bias becomes quite apparent above $10^{13}$ with higher mass
found in the reconstruction. The
histograms make the high mass bias clearly apparent in
\figref{fig:md_300Ac1}, with 90\% pairs in the higher mass bins being
found with a higher mass in the reconstruction, and 20 \% with twice
the mass in the reconstruction. As we explore higher HMC smoothing lengths with
Figures \ref{fig:md_300Ac2} and \ref{fig:md_300Ac3}, the
  distributions of points
  appear centered closer to the equal mass
  diagonal represented in purple. This is confirmed by the black histograms and the
  probability of finding mass excess of 50\% and 100\% appear similar
  in both
  the original and the reconstructed simulations.

As \citet{WangMoYang2014} suggest, for a structure formation model
that is valid on a scale larger than $\sigma_{\rm HMC}$, the HMC
method will introduce non-Gaussian perturbations (non-Gaussianities)
into the reconstructed initial conditions. These non-Gaussian initial
conditions eventually lead to such an excess in halo masses. As we further
confirm here, increasing the
smoothing length significantly reduces the mass bias. However it reduces the mass
range within which halos are efficiently reconstructed. But
  we only consider a reconstruction model with $\sigma_{\rm PM} = 1.5$
  Mpc/h and N$_{\rm PM}=10$. As mentioned in \citet{WangMoYang2014},
  to avoid the mass bias the smoothing scale should be of the order of
  3 PM cells. In order to improve the mass range  with smaller $\sigma_{\rm HMC}$
  smoothing, the $\sigma_{\rm PM}$ parameter must be
  decreased at the cost of increasing the number of steps N$_{\rm
    PM}$. This was demonstrated by \citet{WangMoYang2014}, with more accurate reconstructions using N$_{\rm
    PM}=40$ and $\sigma_{\rm PM}=0.75$ Mpc/h and  $\sigma_{\rm
    HMC}=2.25$ Mpc/h. Such reconstruction models (with low
  $\sigma_{\rm PM}$ and high N$_{\rm
    PM}$) are more
  computationally expensive. They are well suited to produce a single
  set of reconstructed initial conditions. Less accurate reconstruction models such as
  the one used here are better adapted for exploration and testing.

\begin{figure*}
   \centering
   \subfloat[$\sigma_{\rm HMC}=2.25$\label{fig:merit_test_Ac1}]{%
     \includegraphics[width=\twocolfigwidth]{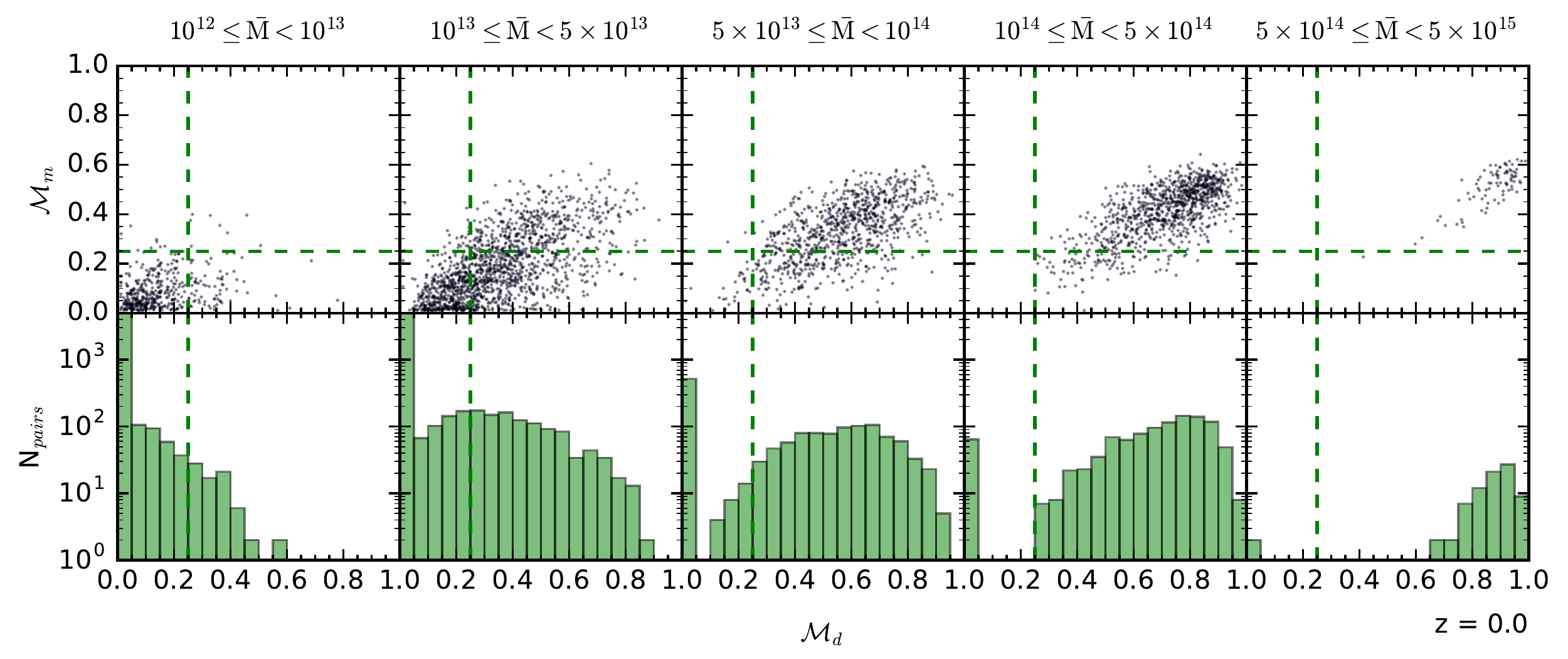}%/merit_test_L300A_vs_L300Ac1_060.pdf}
   } \\
   \subfloat[$\sigma_{\rm HMC}=3$\label{fig:merit_test_Ac2}]{%
     \includegraphics[width=\twocolfigwidth]{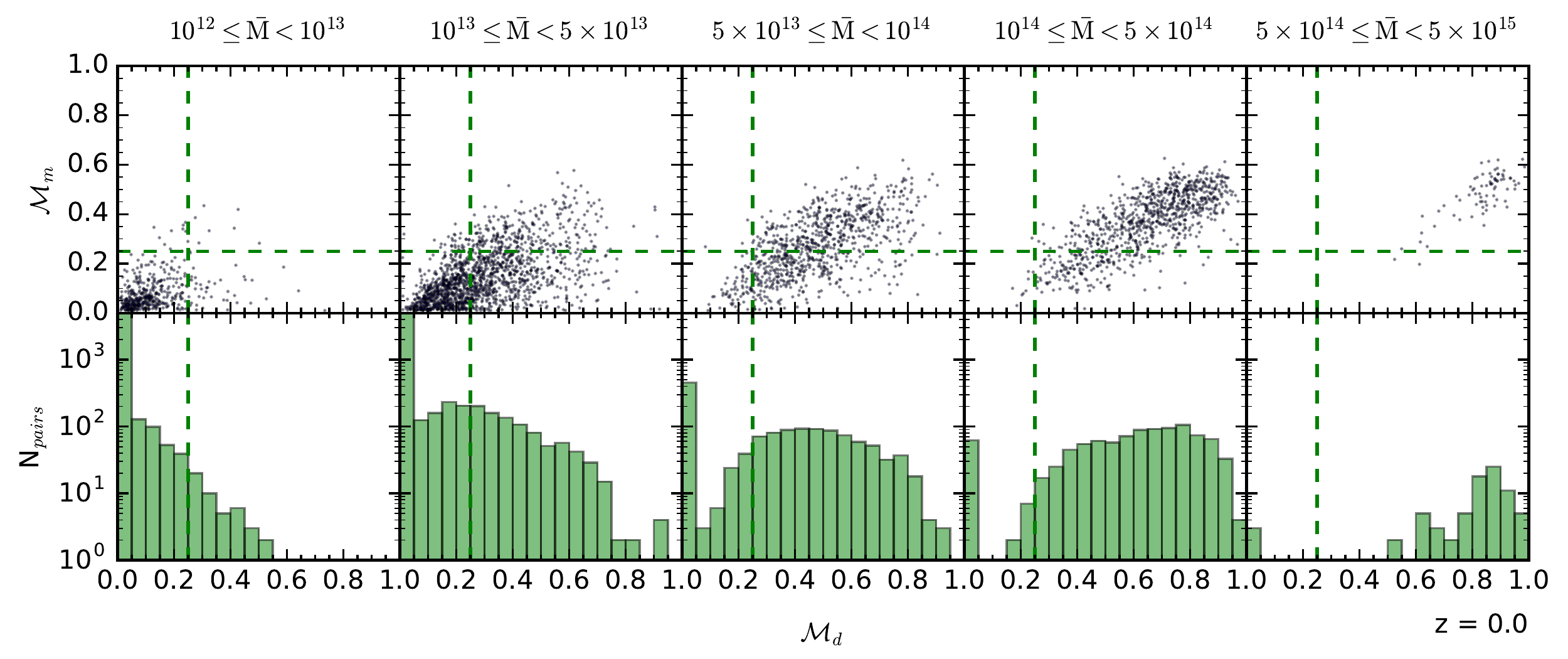}%/merit_test_L300A_vs_L300Ac2_060.pdf}
   } \\
   \subfloat[$\sigma_{\rm HMC}=4.55$\label{fig:merit_test_Ac3}]{%
      \includegraphics[width=\twocolfigwidth]{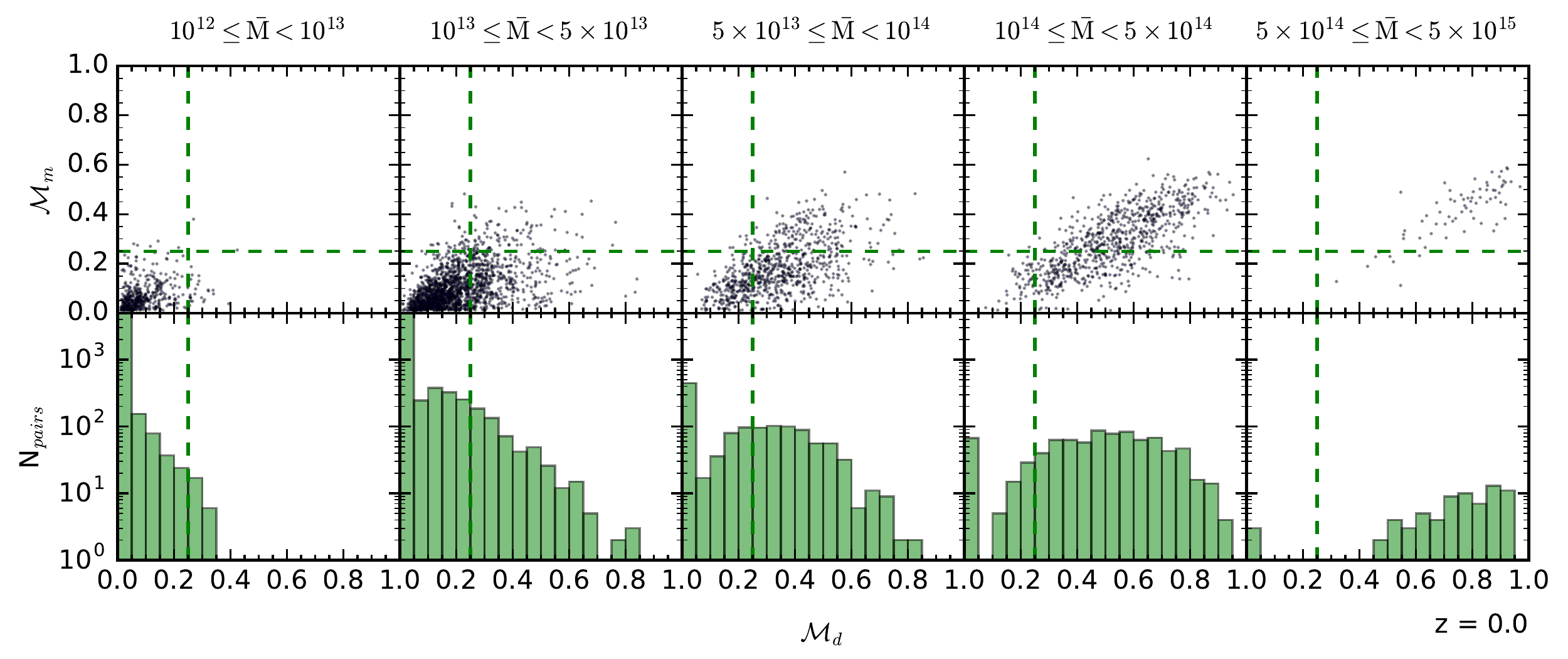}%/merit_test_L300A_vs_L300Ac3_060.pdf}
    } \\
  \caption{Each subfigure assesses the performances
      of one of the reconstruction of the same original
      simulation. For each halo pair cross-identified using our mass based
      criteria (\equaref{eq:merit}) we estimate a distance based merit
      (abscissa, \equaref{eq:merit_dist}). The top panels show the scatter of
      the former as a function of the latter. The bottom panels display the number
    distribution of the cross-identified pairs as a function of the
    distance based merit. Each column corresponds to a
    different mass range for median halo mass. As a guide we indicate
    as dashed green line merit thresholds of 0.25.\label{fig:merit_test}}
\end{figure*}

\subsection{Testing alternative merit functions}
\label{sec:testth}

So far we have implemented a simple merit function as a criteria for
cross-identification. We know that our implementation depends solely
on tracing mass from the initial matter distribution. However, the
objective of the ELUCID project is to reconstruct halos at the same
positions. In order to quantify how close cross-identified halos are,
we define this additional distance based merit function:
\begin{equation}
 \mathcal{M}_d(A,B) = \left ( \frac{r_{\rm FOF,A}+r_{\rm
       FOF,B}}{\left ( r_{\rm FOF,A}+r_{\rm FOF,B} \right ) +\left | \vec{p_A}- \vec{p_B}\right | } \right )^2 \,,
%1 - \frac{3}{2}\frac{\left | \vec{p_A}- \vec{p_B}\right |}{\left
 %     | \vec{p_A}- \vec{p_B}\right | +\left(r_{\rm FOF,A}+r_{\rm FOF,B}\right)} \,,
  \label{eq:merit_dist}
\end{equation}
where $\vec{p_A}$ and $\vec{p_B}$ are the positions of cross
identified halos A and B and $r_{\rm FOF,A}$ and $r_{\rm FOF,B}$
their radii.
We
note that for simplicity, halo positions are defined by their center of
mass. Their size $r_{\rm FOF}$ is estimated by the
maximum distance of their particle to the center. Effectively, the
halos are described as the smallest sphere containing all their
particles. One can also note that, for measuring the distance between halos
$\left |\vec{p_A}- \vec{p_B}\right |$, one needs to take periodic
boundary conditions into account.  The distance criteria we obtain with
\equaref{eq:merit_dist}, is fairly simple, the value goes
down to 0 as
halos are further from one another and rises to 1 if they are
closer. If the halos overlap, $\left | \vec{p_A}- \vec{p_B}\right | <
r_{\rm FOF,A}+r_{\rm FOF,B} +\left | \vec{p_A}- \vec{p_B}\right |$,
then the result of this merit function is larger than 0.25.

We compare directly in \figref{fig:merit_test} (top panels) the two
merit functions for different mass bins (using the median mass of the
cross-identified pairs). The cross-identified catalog used to create this figure differs from
 those used in Figures \ref{fig:mf_smooth} and  \ref{fig:md_smooth}. We
 have not
 used any merit threshold, in order to obtain the equivalent distance
 based merit for unlikely cross-identified pairs. The vertical and
 horizontal dashed lines display the value of our fiducial merit
 threshold of 0.25.
The bottom panels display the distributions of
pairs as a function of the distance based merit. We explore in Figures
\ref{fig:merit_test_Ac1}
to \ref{fig:merit_test_Ac3} the different reconstructions of
simulation L300A.  For all reconstructions, we clearly see in the top
panels, that low mass cross-identified halos do not appear to be close in
the simulation volume and typically have low values in
the corresponding mass derived merit function. For
  reconstruction, as we look at
increasing mass bins, a new population with distance based merit larger than 0.25
emerges. This
bi-modality is made even clearer as we focus on the bottom panels
which reveals the number distributions of cross-identified pairs as a
function of distance based merit.

These figures show that high mass halos tends to be reproduced
with high mass based merits. And as we explore those higher masses,
these pairs tend to be quantified with high distance based
merits. The two merits functions are strongly correlated for halos
pairs that have been most probably correctly cross-identified, especially above $5\times10^{13}$--$10^{14}$ \msun.

\begin{figure*}
  \centering
  \subfloat[$\sigma_{\rm HMC}=2.25$\label{fig:cummf_merits_Ac1}]{
    \includegraphics[width=\twocolfigwidth]{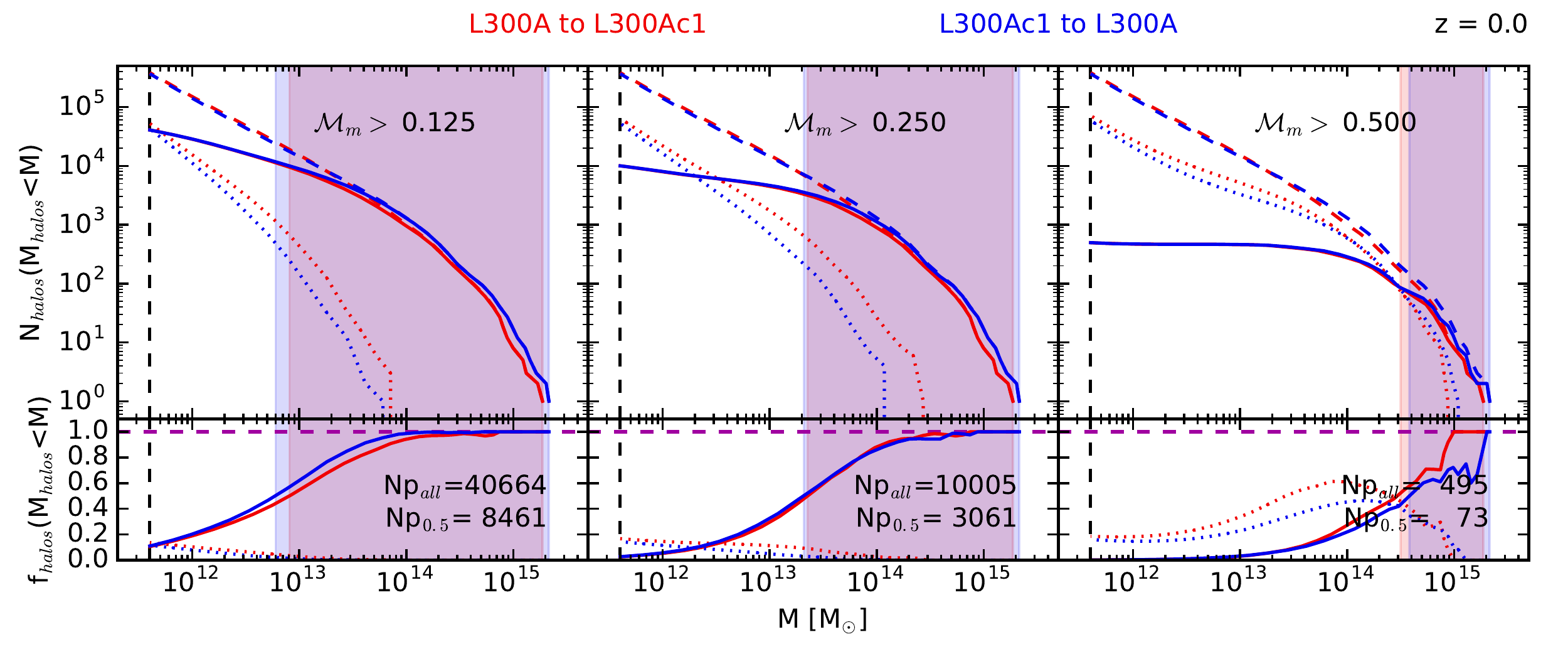}%/cummf_merits_L300A_vs_L300Ac1_060.pdf}
  }\\
  \subfloat[$\sigma_{\rm HMC}=3$\label{fig:cummf_merits_Ac2}]{
    \includegraphics[width=\twocolfigwidth]{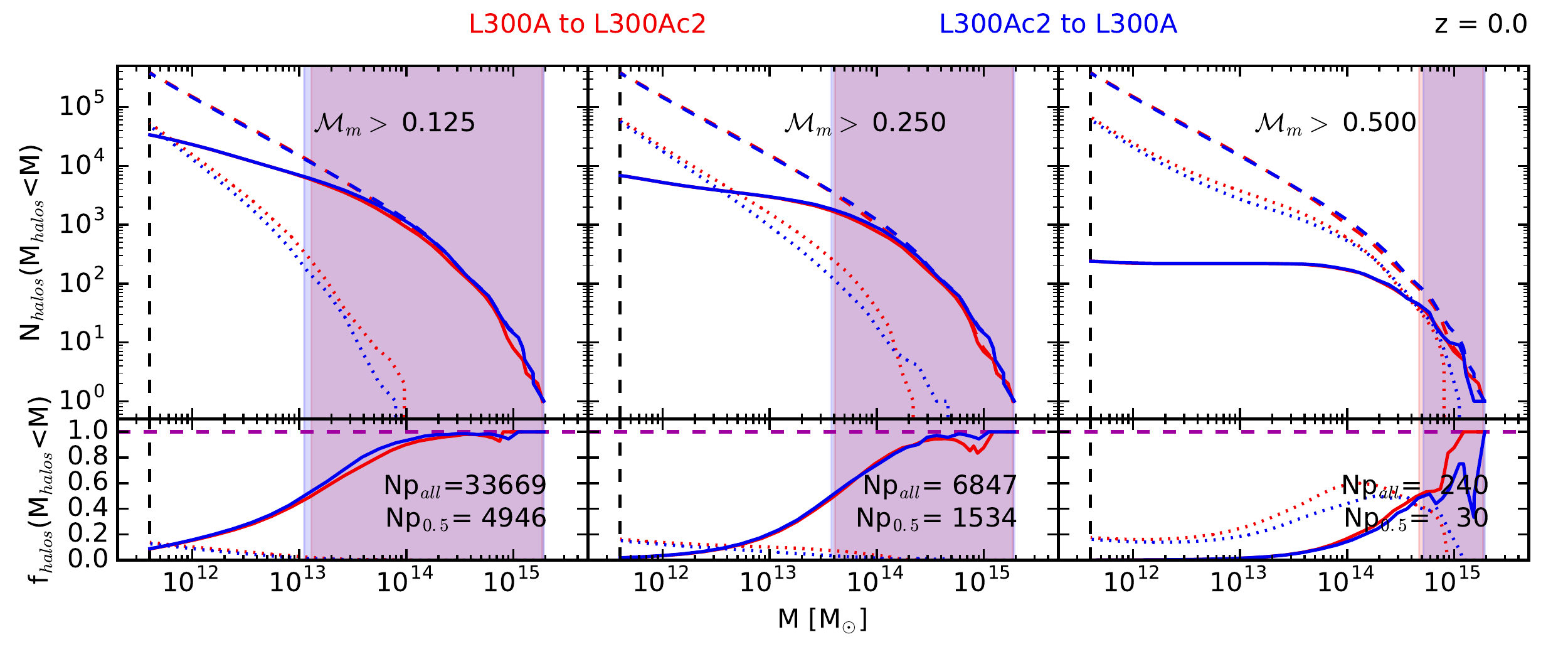}%/cummf_merits_L300A_vs_L300Ac2_060.pdf}
  }\\
  \subfloat[$\sigma_{\rm HMC}=4.55$\label{fig:cummf_merits_Ac3}]{
    \includegraphics[width=\twocolfigwidth]{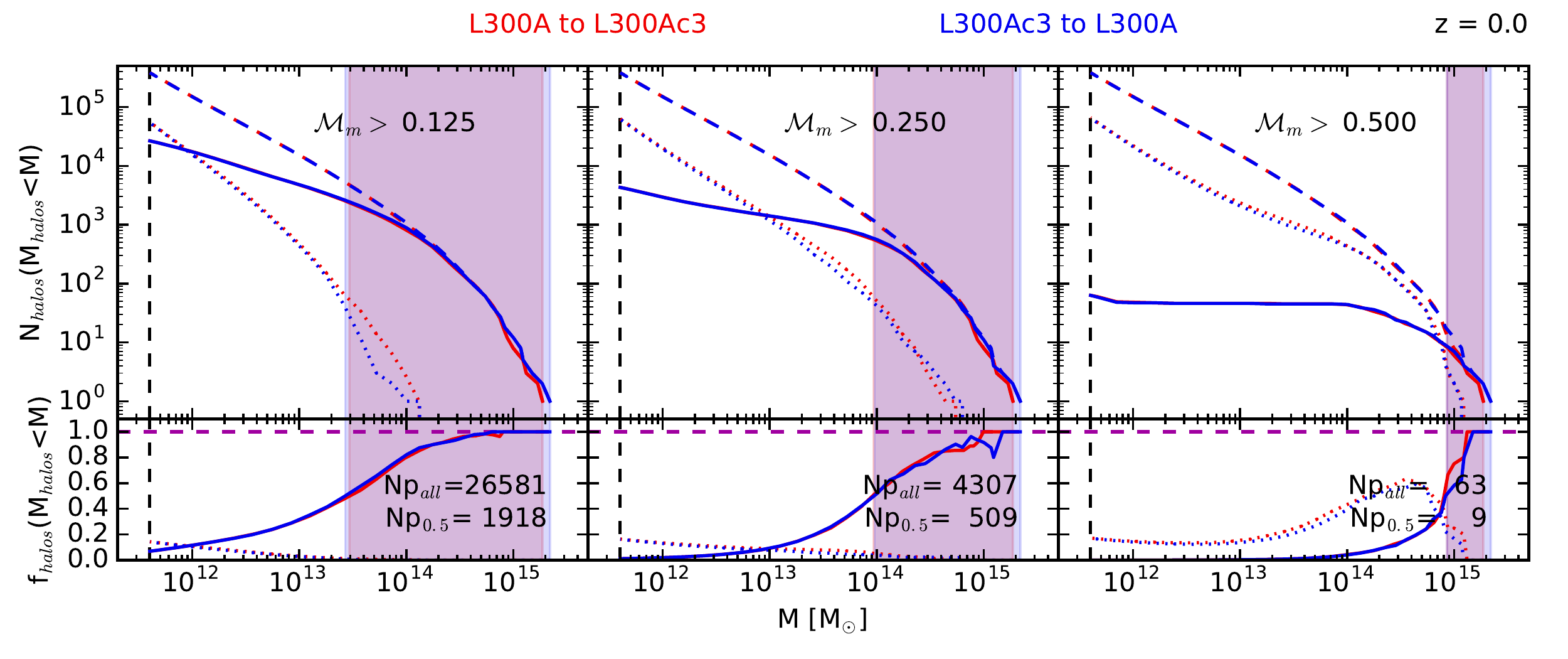}%/cummf_merits_L300A_vs_L300Ac3_060.pdf}
  }
  \caption{Each subfigure assesses the performances
      of one of the reconstruction of the same original simulation.
    Each column contains panels similar to Figures \ref{fig:mf_300Ac1} to
    \ref{fig:mf_300Ac3}. For each subfigure, each column
    correspond to a different selection threshold applied on the merit. Increasing the merit cut reduce the amount of halos pairs
    in the cross-identified catalog as well as mass range for which
    the reconstruction can be considered 50\% complete.
    \label{fig:cummf_merits_A}}
\end{figure*}

For any of the smoothing length, we find that these figure are
very useful for merit function calibration. For the distance based
merit function, the value of 0.25 corresponds to overlap. It is
represented by the vertical green dashed line. For our fiducial merit (mass based)
function, we find a very low number of non-overlapping
cross-identified pairs above our fiducial 0.25 threshold (horizontal green dashed line)
. This result shows that we can indeed apply this mass based
cross-identification approach to such simulation reconstruction
problems.

We can now decide to use  our fiducial mass based merit function as a
  baseline. We suppose that we wish to
  reproduce the predictions from the mass based merit using the
  distance based merit. The green dashed lines divide the top panels
  into 4 domains. We find the
  true negatives and true positives in the lower left and upper right
  corners respectively. In the upper left and lower right, we find the false
  negatives and false positives respectively. We can see that the number of false
  negatives is very low  compared to the number of false positives. This
 suggests that the performances of the distance based
  cross-identification may be improved by using a larger merit
  threshold.

 We have explored an existing cross-identified sample to test
  a potential distance based merit function. The data we used here maximize the value of the
  mass based one. The distance based merits shown here are
  over-estimated compared to what would be found by a cross-identification algorithm
  using this criteria.
 One must keep in mind that this
merit function cannot be fully tested unless implemented in a
cross-identification algorithm.

\begin{figure*}
  \centering
  \subfloat[$\sigma_{\rm HMC}=2.25$\label{fig:cummf_merits_Bc1}]{
    \includegraphics[width=\twocolfigwidth]{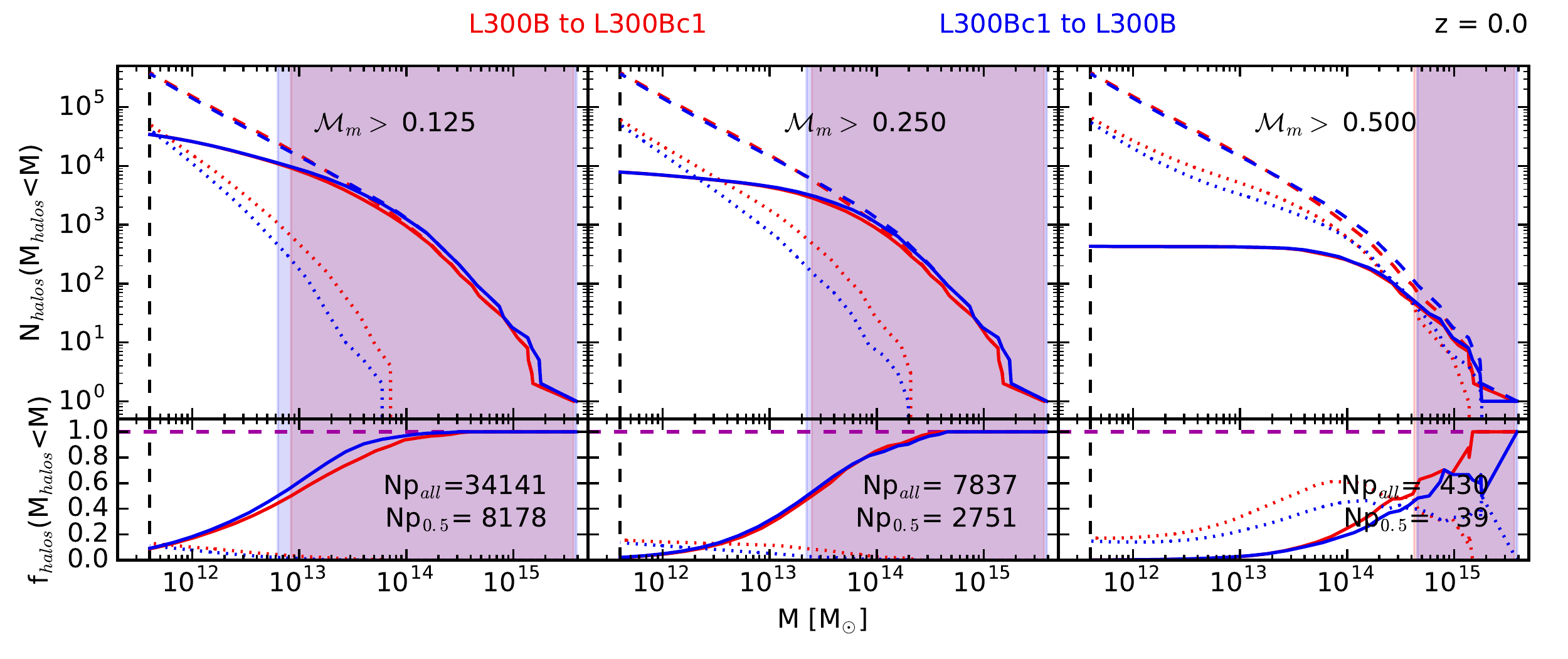}%/cummf_merits_L300B_vs_L300Bc1_060.pdf}
  }\\
  \subfloat[$\sigma_{\rm HMC}=3$\label{fig:cummf_merits_Bc2}]{
    \includegraphics[width=\twocolfigwidth]{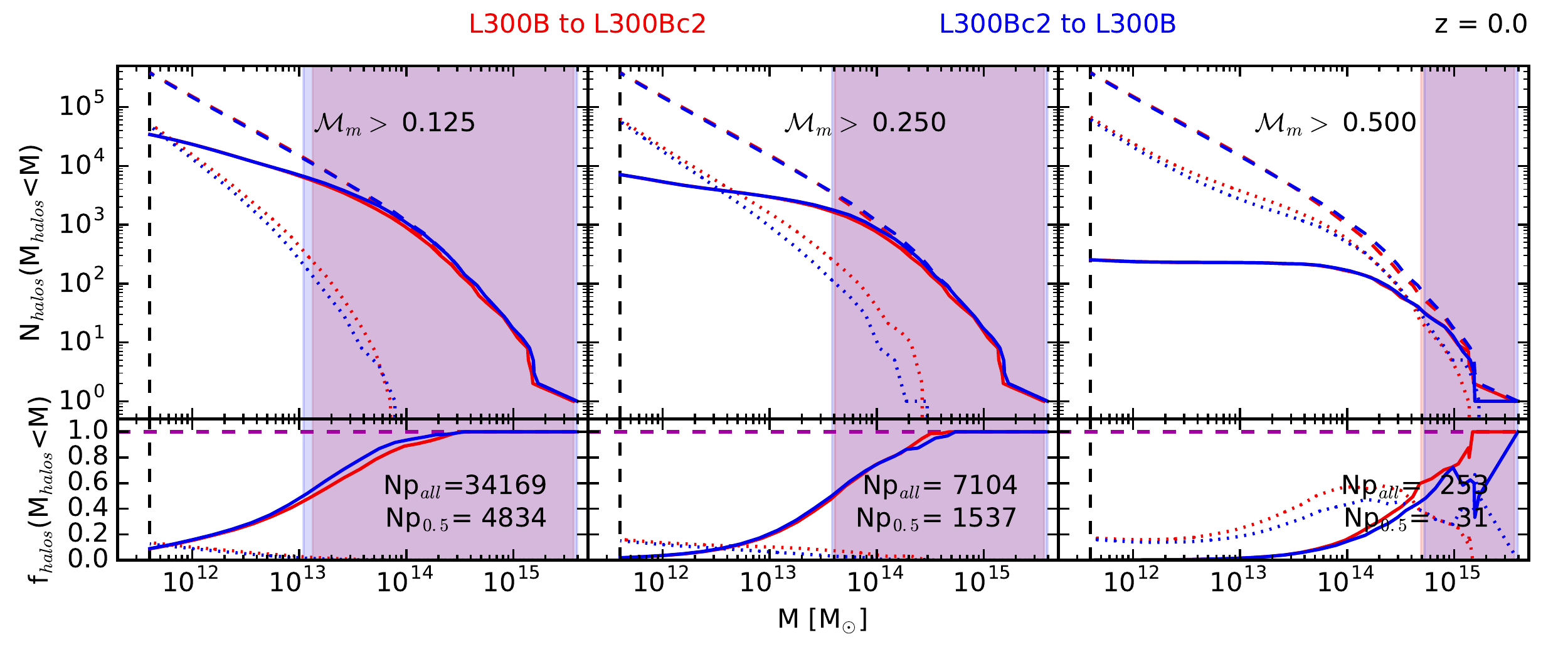}%/cummf_merits_L300B_vs_L300Bc2_060.pdf}
  }\\
  \subfloat[$\sigma_{\rm HMC}=4.55$\label{fig:cummf_merits_Bc3}]{
    \includegraphics[width=\twocolfigwidth]{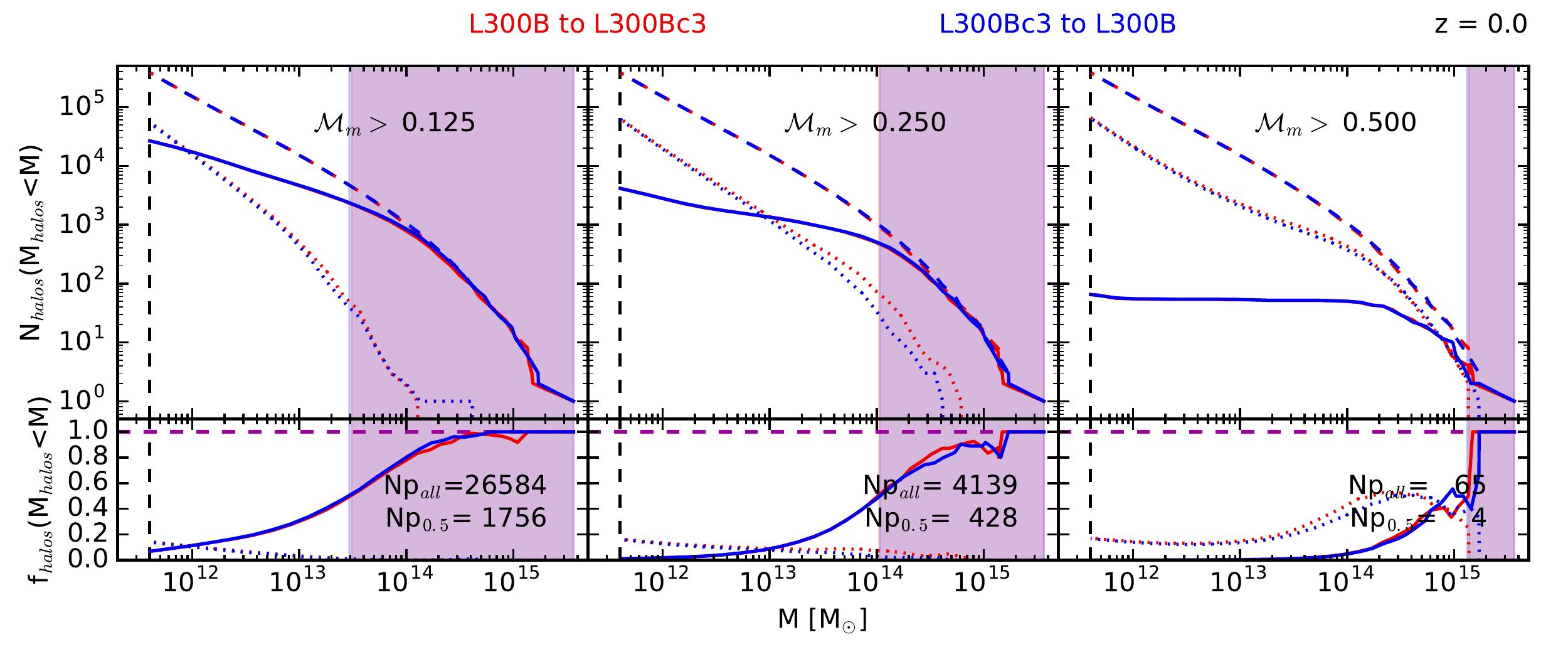}%/cummf_merits_L300B_vs_L300Bc3_060.pdf}
  }
  \caption{Same as \figref{fig:cummf_merits_A} with the L300B
    simulation and its reconstructions. Results are qualitatively the
    same but we find a strong
    variation (up to 25 \%)  in terms of  number of
    cross-identified pairs for similar sets.
    \label{fig:cummf_merits_B}}
\end{figure*}

\subsection{Applying different merit thresholds}
\label{sec:Mthreshmf}

In \figref{fig:cummf_merits_A}, we explored the cumulative mass functions
after cross-identification with multiple merit thresholds, for
the different reconstructions as subfigures \ref{fig:cummf_merits_Ac1} to
\ref{fig:cummf_merits_Ac3}.
The columns correspond to different values of the merit
  threshold, increasing from left to right.
Our fiducial
value of 0.25 used in the previous Figures \ref{fig:mf_300Ac1} to
\ref{fig:mf_300Ac3} are displayed in the central panel.

We note that this selection threshold described and implemented
here is not a parameter of the cross-identification method described
in \pararef{sec:alcomp}, but is applied a posteriori to generate cross identified
and associated catalogs. Each member of any
halo pair that does not meet the merit threshold is
  added to the associated catalog provided that 50\%
of their mass is found in the other member.

\begin{figure*}
  \centering
  \includegraphics[width=\twocolfigwidth]{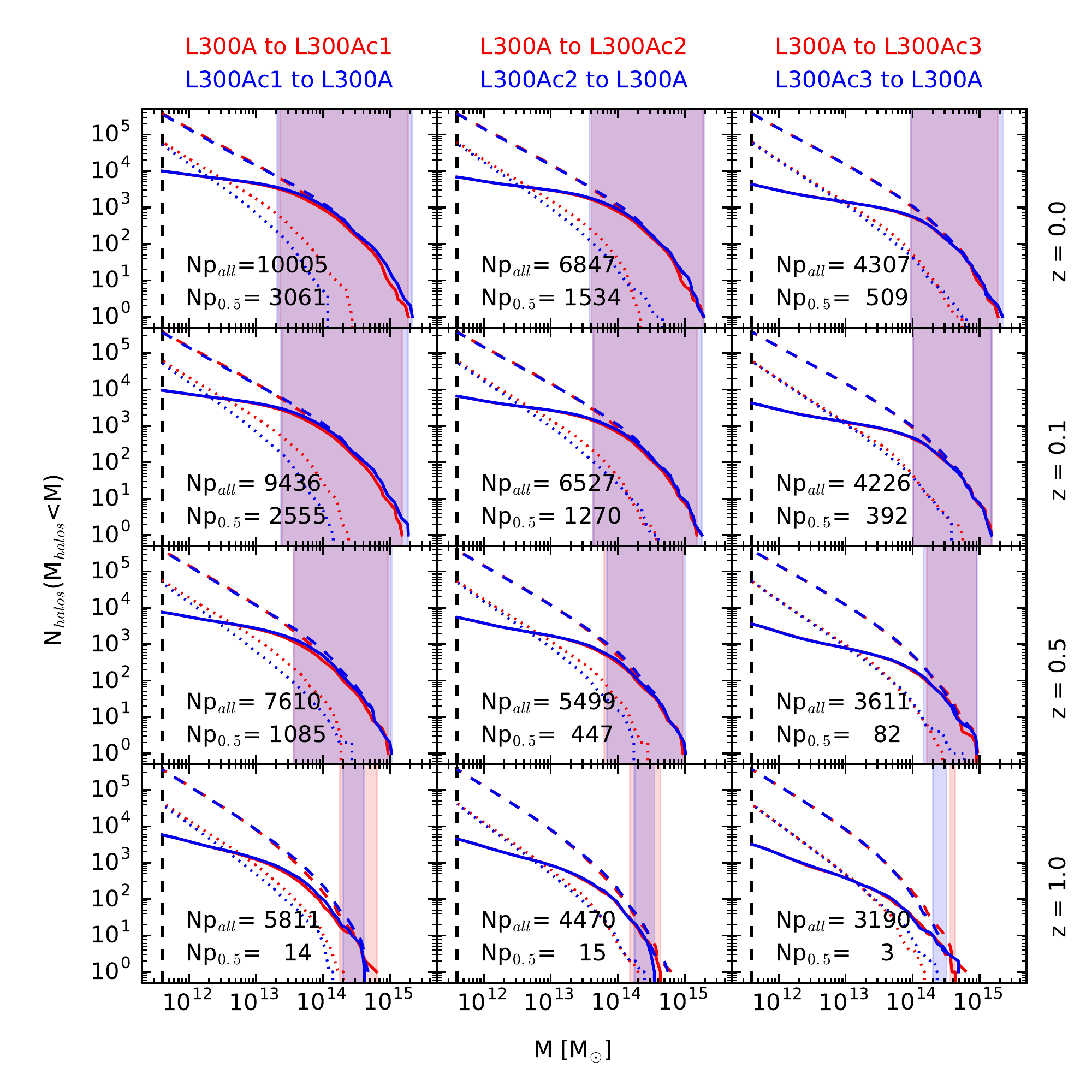}%/cummf_z_L300A_vs_all.pdf}
  \caption{Similarly to the upper panels of Figures \ref{fig:mf_300Ac1}
  to \ref{fig:mf_300Ac3}, we assess the reconstructions through the
  cumulative mass functions. Each column corresponds to a different
  reconstruction of the same original simulation, each row correspond
  to a different redshifts. As we explore larger redshift
    the reconstruction is degraded both in term of number of
    cross-identified pairs but also in term of the 50\% completeness mass range.
    \label{fig:mf_allz}}
\end{figure*}

As expected, reducing the value of the merit threshold increases the number of halos to be considered cross-identified.
Compared to our fiducial value, dividing this selection threshold by
2, increases the number of halos in the cross-identified catalog by a
factor 4 to 6  depending on the reconstruction  (HMC
smoothing). We have obtained 27000 to
34000 paired halos with this lower threshold, instead of 4000
to 8000 pairs with the fiducial value of 0.25. Allowing for a
smaller merit strongly increases the size of the subsample that
can be used for cross-comparison. The downside is that a lower value in
terms of selection would allow for a much larger mass difference
between cross-identified halos. Moreover as
\figref{fig:merit_test} suggests, halo pairs with low merits
may also be located at larger distances within the simulation
box. As we double the value of our
fiducial selection threshold, the size of the
cross-identified sample is greatly reduced, by a factor of 30 for the
smallest smoothing length to a factor of 65 for the largest one.
Despite its small size, this catalog should have the advantage
  of showing halo pairs of very similar masses at close coordinates.

The 50 \% completeness region is also strongly affected, with its lower limit moving toward the high mass end (upper limit)
with increasing merit threshold. The mass range discrepancy
is also more apparent as we consider lower merit thresholds. The merit
is thus a very effective criteria to create either a small but
accurate reconstructed subsample or a very large but much less
  accurate one. The former may be preferred for very detailed analysis
  of specific halos while the latter could be more
relevant for large number statistics.

Within the framework of comparing reconstructed simulations, our
fiducial value seems to be a good compromise. The size of
cross-identified catalog is still dependent on the
realization itself. As we go through the same exercise with the second
set of simulations L300B in \figref{fig:cummf_merits_B}. For the
smallest value of $\sigma_{\rm HMC}$, we found 15 to 25 \% less
cross-identified pairs than in the L300A simulation set. There
is also a
10\% discrepancy in the opposite direction for the medium value of
$\sigma_{\rm HMC}$. The difference is less apparent for the largest
smoothing, unless we focus on the 50\% completeness region where we
find larger numbers in the L300A simulation set. This remark
highlights the fact that for a specific set of parameters used in the reconstruction the
number of reconstructed halos may differ to some potentially
large extent. However one aspect of the reconstruction which does not
seem to be affected is the 50\% completeness mass range. This suggests
that the mass range for which we obtain a certain level
of completeness is a better indicator for testing a new set of
reconstruction parameters.

\begin{figure*}
  \centering
  \includegraphics[width=\twocolfigwidth]{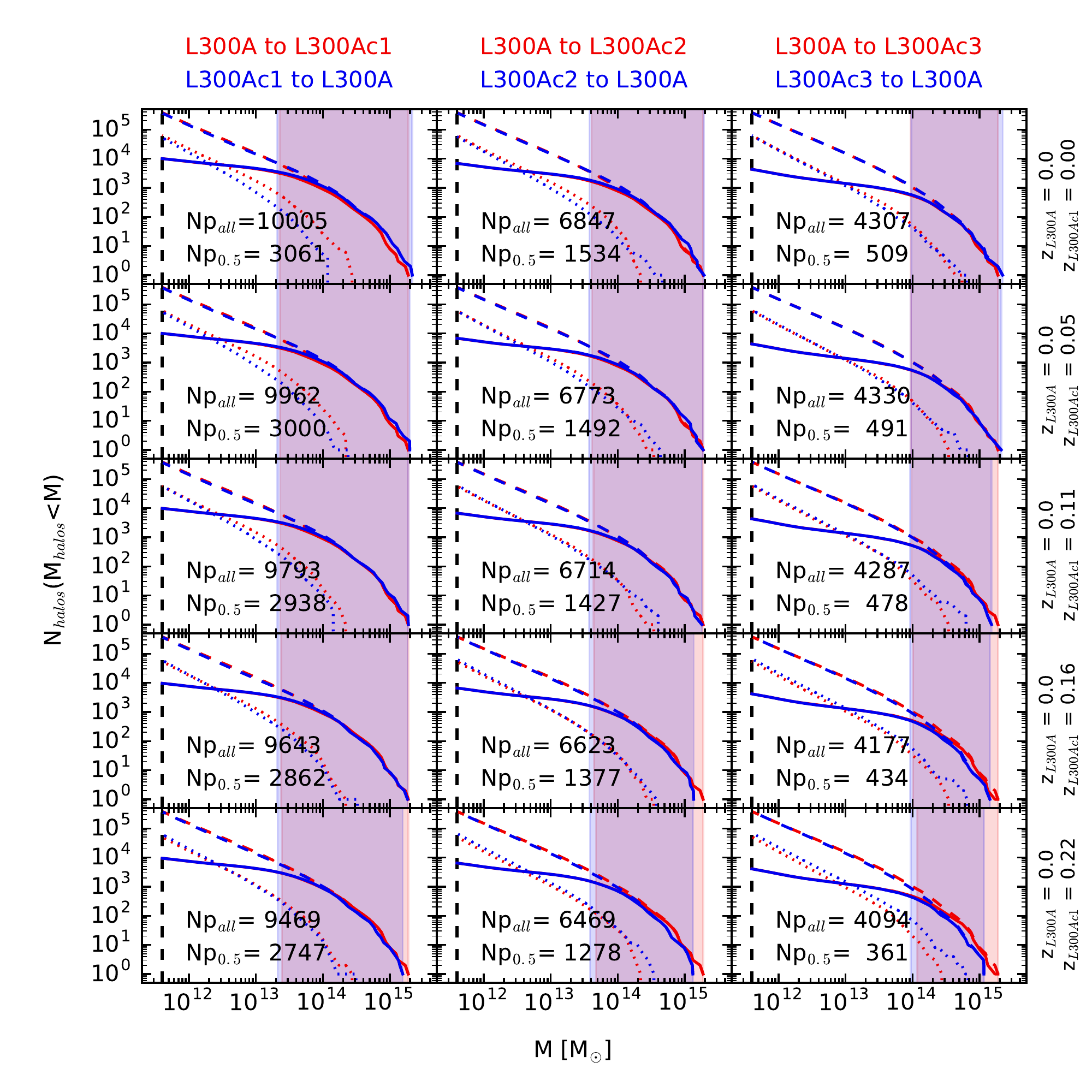}%/cummf_L300A_vs_all_Deltaz_059.pdf}
   \caption{This figure is similar to
       \figref{fig:mf_allz}. Instead of exploring earlier redshifts,
       we compare earlier snapshots of the reconstruction to the final
       output of the original simulation. Even though
     earlier reconstructed mass functions are a better match
       to the final original mass functions, the cross-identified
     catalogs contain less pairs, and the 50\% completeness mass range
     gets smaller with increasing time difference.
     \label{fig:mf_all_Deltaz}}
\end{figure*}

\subsection{Cross-comparison at earlier redshifts}
\label{sec:compz}

So far, we have shown that the z=0 output is efficiently
reconstructed at the higher end of the mass function. This result is
obviously by design since reconstruction constraints come from the
z=0 smoothed density map.
Given that the larger structures should also be present in the
simulation, we proceed to compare earlier snapshots to find
  out whether, by construction, we obtain some constraints at
earlier times.

The most logical step to begin with is to directly compare the
mass functions for earlier redshifts. In \figref{fig:mf_allz}, we run the same test as
in \pararef{sec:smooth} but at multiple redshifts. Each column corresponds
to a different reconstruction of the same original simulation,
  with a different smoothing scale $\sigma_{\rm HMC}$ applied to the
density map. Each row corresponds to the different redshifts 0., 0.1,
0.5 and 1 respectively from top to bottom. As in
Figures \ref{fig:mf_300Ac1} to \ref{fig:mf_300Ac3}, we display the
number of cross-identified halo pairs (with merit above 0.25) and the
mass range corresponding to 50\% completeness, the corresponding
numbers are also displayed.

Independently of the smoothing length used, one can see in this figure
how the reconstruction gets degraded as we explore larger
redshifts. The minimum mass
from which half the halos should be reconstructed is higher at larger
redshift. Notably the number of halos reconstructed above that mass
gets smaller as well. If the growth of halos was purely linear, we
would expect the high mass end to be populated by the same
  halos taken at different times and the 50 \%
  completeness mass range would simply shift to the left.

Cross-identified halos at redshift 0 most certainly have
  multiple progenitors at earlier times. Their main progenitors may also be
  cross-identified but their secondary progenitors are much
  less likely to be
constrained by the reconstruction technique. The
  contribution of these secondary progenitors to the mass function
  with the degradation of the cross-identification of the main
  progenitor explain why the minimum of the shaded region
  increases.  The cross-identification gets so degraded
 by redshift 1 that the most massive halos are no longer
 cross-identified. It is especially evident for the largest smoothing
 lengths as the 50 \% completeness regions no longer
  overlap.

This degradation can be understood as variation of the merger
  histories of specific halos. For example, we could imagine a perfectly
  cross identified pair. They have the same particles, translating to
  a merit of 1. As we explore earlier time outputs, we can imagine that the original
  halo is split in two parts, the reconstructed halo being unchanged. The larger part would still be
  cross-identified but with a lower merit, the smaller part would be
  associated. At an even earlier time output, it is the reconstructed
  halo that is split. But a different subset of particles could be removed
  (the split fraction could be
  identical). The merit of the cross-identified pair would once again be
  lowered and an additional associated halo found in the reconstruction. As we imagine
that each split is a new progenitor, we understand how individual
progenitors (including the main) may no longer be
cross-identified. However, the collection of progenitors of the
original halo  could be cross-identified with the
collection of progenitors of the reconstructed one.

\subsection{Cross-comparison across redshifts}
\label{sec:compzdz}

We pointed out in \pararef{sec:smooth}, that in the high mass end of
the mass function, we find more massive
halos in the reconstructed simulation. The issue may be related to an over-merging
process, with nearby original halos being reconstructed as
one. The halos may thus appear
more evolved than in the original simulation. We investigate this
possibility in \figref{fig:mf_all_Deltaz}. It is similar to
\figref{fig:mf_allz} with one key difference, the redshift of the
original simulation remains fixed (z=0), but we use
5 consecutive snapshots of the
reconstructed simulations.

The mass functions of the
reconstructions appear to be closer to the original mass
functions if we use an earlier time step in the reconstruction, respectively z=0.16 for L300Ac1, z=0.11 for L300Ac2 and
z=0.05 for L300Ac3. We notice that as the high mass issue is less relevant for
the given reconstruction (larger smoothing length) we naturally pick
out a closer time output. The cumulative mass function of cross-identified
halos follows the same trend as the mass function, and the
discrepancies between the associated mass functions get mitigated
at earlier time steps in the reconstruction.

These qualitative remarks tend to suggest that the
  reconstructions are more evolved than the original. If the
  reconstructed halos are mergers of a cross-identified associated
  original halos, we could find an earlier time step in the reconstruction where the mergers
  have not yet occurred. In this case the main reconstructed halos would
  have no mass excess, and the associated original halo would now be
  cross-identified. Then the number of cross-identified pairs
  should increase, and the lower limit of the 50\% completeness
  region decrease (at least slightly).

Once we focus on the numbers of reconstructed halos, we notice that
the performances of the reconstruction does not appear to be improved. The number
of reconstructed halos in the original simulation decreases. We can
also note some variation in the 50\% completeness region, with
the region corresponding to the reconstruction (in blue) being shifted to lower
mass while the one corresponding to the original simulation appears to
remain fixed. Even though the mass functions are closer when we compare different
redshifts. In terms of number of reconstructed halos, the quality of the reconstruction is not significantly
improved. The most massive halos are simply a little less
  massive, and less original halos are associated to the reconstructed
  ones. The reconstruction is thus not more evolved than the
  original simulation.

%==============================================================

\section{Summary and Discussion}
\label{sec:discuss}

This paper can be summarized as follows:
\begin{itemize}
\item We have implemented a simple and reliable cross-identification method for halo catalogs derived from $N$-body simulations.
\item We have used the cumulative mass
functions to qualify the comparison and the
numbers of cross-identified pairs and 50\% completeness mass ranges to
quantify it.
\item We have applied this method on simulations
    obtained by
    \citet{WangMoYang2014}  to test the initial condition
  reconstruction method used for the ELUCID project.
\item Our cross-identified catalogs have been further tested with
  a distance based criteria. The results are highly compatible at the high
  mass end where 50\% of the halos are reconstructed.  With
    our default selection threshold, a very limited number of
  non-overlapping halos are part of the cross-identified catalogs.
\item We have confirmed the effect of the smoothing parameter
  $\sigma_{\rm HMC}$ on the reconstructed halo catalog. We
    have also highlighted how a large
  value could avoid an artificial increase of the halo mass at the high end of
  the mass function to the detriment of the quantity of
  reconstructed halos.
\item Even for the same set of reconstruction parameters, the
  quantity of reconstructed halos can vary between
  realizations.
 The reconstructed (50\% completeness) mass range might be a reliable indicator.
\item We have further explored the quality of the
  reconstruction and have confirmed
  a degradation of the results at higher redshifts.
\end{itemize}

The whole scope of the ELUCID project is to produce reliable initial
conditions that could lead to more accurate reconstructions of the large
scale structures as observed in SDSS. Applying reconstruction methods
to cosmological simulations with a large number of constraints is not
trivial. Such simulation tests are extremely useful as they can be used
to benchmark the reconstruction algorithms, weighting the efficiency of
the code itself with the quality of the reconstruction.

The diagnostic scheme we implemented in this paper differs considerably from
that of \citet{WangMoYang2014}, as it focuses on the end
result of the simulations by one-to-one matching. We have
 confirmed their observation of the mass bias in more
  details. Furthermore, by quantifying the reconstruction, we have also ruled out the possibility that
this bias was due to a faster evolution in the reconstructed simulations.
The
cross-identified catalog is another tool to diagnose discrepancies once
either a HOD (Halo Occupation Distribution) or a SAM
  (Semi-Analytical Model) are applied to
the simulations. Systematic differences between mock catalogs should be directly
assessed in light of the reconstructed halo properties themselves.

An additional aspect that may be relevant to such a study is
the impact of the group-finder. The friend of friend (FOF) algorithm is one
of the simplest and the most widely used one, even as an
initial guess for more accurate and reliable algorithms. The high mass
end discrepancies we observed may become less apparent when using
alternatives. Applying a subhalo finder on top of the FOF halos could
separate coalesced halos for the other simulations. Still given the
smoothing, one should not expect subhalo distributions to be a close
match for reconstructed halo both in term of mass and in term of
position. Defining halos as bound spheres using a SOD like algorithm
\citep[Spherical Over Density;][]{LaceyCole1994}, could also mitigate the problem.
However neglecting particles either unbound  or outside the virial radius
could reduce the chance of cross-identifying halos thus introducing a
bias. We preferred the simplest method for this analysis. In
  order to identify differences between
 original and reconstructed mock galaxies, one should
   cross-identify the (sub)halo catalogs that are part of the semi-analytical pipeline.

An extremely interesting aspect of this study is not
  only the cross-identification itself but also the nature of the cross-identification criteria
The merger-tree structure is indeed useful for
catalog comparison.
Since a classic a merger-tree code uses the particle indexes to find
connections between halos. One can see why tree-builders have been used efficiently to compare group catalogs extracted from the same
simulation, for example in confronting two group-finders.
Such method are useful for studying the effect of baryons in simulations (as in
  \citet[as in][]{Cui2012, Cui2014}, as it requires to
  cross-identify halos from the hydrodynamical run with the $N$-body one.
A further application of such a method could be resolution study. Taking
the same initial conditions applied at 2 resolutions (e.g. a factor 8 in
particle number),  one can map the
correspondence of particle indexes between simulations and
then modify a tree-builder for halo cross-identification
  across resolutions.

In our case we used two sets of initial conditions. Even if
halos could be produced with the same mass in the same positions,
there is no apparent reason why they would contain the same particle indexes. However, since particles are effectively indexed by their initial Lagrangian grid
coordinates, they are not random. That leads one to
conclude that our method relies on the fact that
cross-identified halos are formed out of the same Lagrangian
regions. This  is quite interesting as it is
actually the concept behind initial conditions reconstruction
techniques.

This brings us to another interesting aspect of this cross-identification
scheme. As reconstructed halos are born out of similar Lagrangian
regions, we could expect their progenitors to be born out of the same
regions themselves. A tree-builder would use the same criteria as
  our cross-identification algorithm. If the merit of a
  cross-identified halo pair is high, their respective main
  progenitors are likely to be cross-identified as well.
We have found, however, that any
consistency at some earlier redshift remains limited to a very limited
redshift range. It would be necessary to combine the
  cross-identification catalogs with merger-trees to understand the
  degradation of the cross-identification.
Future study on this line would answer the following
questions,
\begin{inparaenum}[(i)]
\item how far back
halos are effectively reconstructed and
\item how far back the
  collection of
their progenitors are reconstructed.
\end{inparaenum}
Similarities in accretion
  histories of cross-identified pairs should also be assessed with respect to halos of similar masses. This
topic is extremely relevant in determining
indirect constraints on Semi-Analytical galaxy mock catalogs due
to the reconstruction.

%==============================================================

\section*{Acknowledgments}
DT thanks Alexander Knebe and Peter Thomas for useful discussions that
initiated the idea behind the method.
This work is supported by the
973 Program (No. 2015CB857002), national science foundation of China
(grant Nos. 11128306, 11121062, 11233005, 11073017, 11621303),
NCET-11-0879 and the Shanghai Committee of Science and Technology,
China (grant No. 12ZR1452800). HW further acknowledge the support form
grants NSFC 11522324 and NSFC 11421303.
 We also thank the support of a key
laboratory grant from the Office of Science and Technology, Shanghai
Municipal Government (No. 11DZ2260700).
This work is also supported by the High Performance Computing Resource
in the Core Facility for Advanced Research Computing at Shanghai
Astronomical Observatory. DT thanks Richard Tweed and the anonymous
referee for helping to improve
the quality and clarity of this manuscript.

\bibliographystyle{apj}

\bibliography{bibliography}

\begin{appendix}

%==============================================================

\section{ELUCID: historical background}
\label{sec: ELUCID}
We describe a short history of the ELUCID project:
\begin{description}
\setlength\itemsep{0 em}
\item[\citet{Yang2005a}] developed a method to associate galaxy groups
  to halos. This iterative method consists in defining galaxy groups
  in observational surveys using a halo mass estimation algorithm, then
  applying a halo occupation distribution model to estimate a halo
  radius and velocity dispersion, and iteratively refining the group membership with improved
  center and halo radius. The method was benchmarked on mock catalogs
  derived from $N$-body simulations \citep[code: ][]{JingSuto2002}
  and applied to the 2dFGRS survey \citep{Colless2001}.
\item[\citet{Yang2007}] applied the \citet{Yang2005a} method to the
  SDSS DR4 \citep{AdelmanMcCarthy2006} survey. Following publications
  assessed the halo occupation numbers in terms of luminosity \citep{Yang2008}, and the
  derived group luminosity and stellar mass functions \citep{Yang2009}.
\item[\citet{WangMoJing2009}] implemented a Monte Carlo reconstruction
 method and applied it to cosmological $N$-body simulations
  run with the P$^3$M by \citet*{JingSuto2002}. This method was
  applied by \citet{MunozCuartas2011} on the SDSS DR4 \citep{AdelmanMcCarthy2006} without taking redshift distortions into account.
\item[\citet{Yang2012}] applied the \citet{Yang2005a} method to the
  SDSS DR7 \citep{Abazajian2009} survey. The stellar mass evolution of
  both central and satellite galaxies were discussed in a self
  consistent way.
\item[\citet*{WangMoYang2012}] used the
  method of \citet{WangMoJing2009} to reconstruct the cosmic velocity field and the associated tidal field in the
  SDSS DR7 \citep{Abazajian2009} survey volume, including redshift distortions.
\item[\citet*{WangMoYang2013}] combined a Hamiltonian Monte Carlo
  Method \citep[HMC: ][]{Hanson2001, Taylor2008} and a Modified Zel'dovich
  approximation \citep[MZA: ][]{TassevZaldarriaga2012b}. This method was
  applied to a mock catalog of SDSS DR7 galaxy population associated to a halo catalog
  extracted from the millennium simulation \citep{Springel2005}, the
  $N$-body code used is {\tt GADGET2}  \citep{Springel2005}.
\item[\citet*{WangMoYang2014}] further modified the model by
  combining the  HMC
  method and a Particle Mesh (PM) method  widely used in $N$-body
  codes. This new model was confronted with the previous one, by testing on
  $N$-body simulations. This paper constitutes the first publication of
  the ELUCID project.
\item[\citet{WangMoYang2016}] applied the HMC + PM method to SDSS
  DR7 using group catalog derived by \citep{Yang2005a,Yang2007}. This
  publications describes the first large scale, high resolution,
  ELUCID simulation. This simulation was performed on the CPU Node $\pi$ at the Center for High Performance Computing (CHPC) in Shanghai Jiao Tong University.
The simulation ran for 178h (7.4 days) on 128 nodes (2048 cores in
total). Each of the 100 time outputs requires about 865 Go disk space.
\end{description}

\section{Cross-identification: Consistency issue and correction}
\label{sec:cor_comp}

Despite the fact that tree-builders can be efficiently applied
  to the halo cross-identification
problem, some further corrections are necessary to ensure self-consistency.
Going back to the illustration introduced in Figures \ref{fig:comp_step4} and \ref{fig:comp_step2}),
starting from a group of
A and tracing the double tipped arrows, one doesn't end up
  at the same halo. We mentioned this
  possibility in the illustrations; in this appendix we illustrate how
  this issue can occur and how we avoid them.

 These cross-identification mismatches come from limiting associations and
potential cross-identifications to one candidate.
At first glance such error may seem improbable since we use the same
definition in both directions,
but it can be explained through \figref{fig:err_tree}. In this
illustration, we have selected the rightmost two halos
from simulations A and B. In
  the right side,
the 3 possible connections  ($A5$, $B7$),
($A5$, $B8$) and ($A6$, $B8$) are represented with dashed lines and
their respective strength (or merit) are labeled $M1$ , $M2$ and $M3$. We see that if the condition $M1>M2>M3$
is satisfied we obtain the associations described in the
middle panel
(comparing B to A) and
right panel (comparing A to B). Once the same criteria is
applied to define the cross-identifications  we are faced with this inconsistency.

\begin{figure}[t]
  \centering
  \subfloat[Illustration of an asymmetry leading to a
  cross-identification inconsistency \label{fig:err_tree}]{%
    \includegraphics[width=\onecolfigwidth]{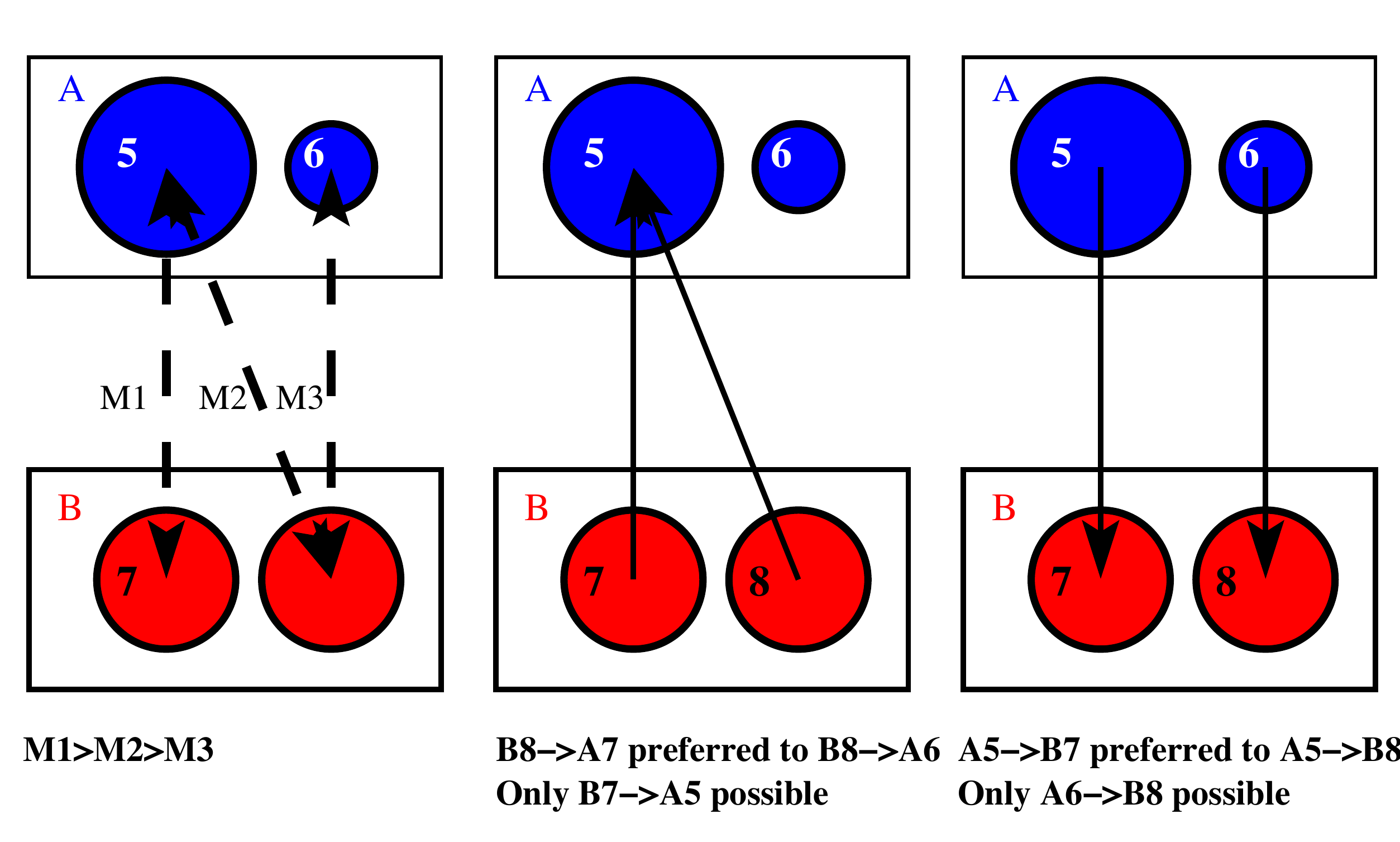}%/err_fig.pdf}
  }
  \subfloat[Result after correction \label{fig:tree_correct}]{%
    \includegraphics[width=\onecolfigwidth]{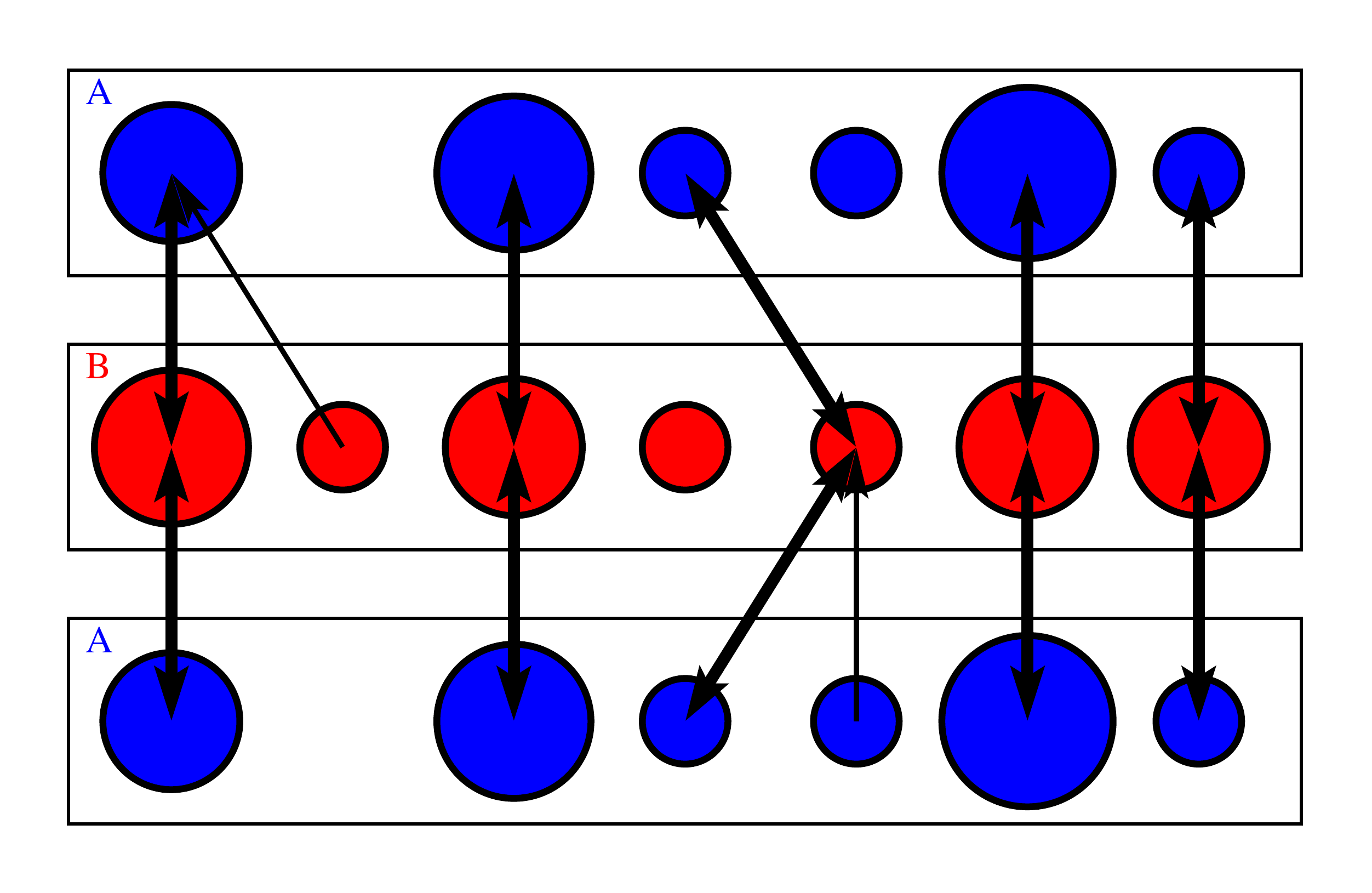}%/comp_fig8.pdf}
  }
  \caption{Illustration of a cross-comparison inconsistency, and
    possible correction. We ensure that the same answers
    are obtained after substituting catalog A and B. The configuration
    illustrated in the left panel is detected in order to obtain the
    final result displayed in the right panel.
    \label{fig:err_correct}}
\end{figure}

Solving it simply requires ignoring at least one
connection. One way to do so is to apply additional conditions
for associating halos, or apply a threshold to the merit.
For a particle based merit function, we could define a criteria that
would render inconsistencies
impossible by construction.
Let us first denote  $f^X_Y = M_{X\cap Y}/M_X$, and suppose our example
has the following characteristics: $f^{B7}_{A5}>0.5$,
$f^{B8}_{A5}>0.5$ and $f^{A6}_{B8}>0.5$.
We have by definition $f^{B8}_{A5}+f^{B8}_{A6}\leq 1$,  so
$f^{B8}_{A6}<0.5$. Imposing the conditions $f^X_Y > 0.5$ and $f^Y_X > 0.5$, would remove
the (A6,B8) connection and the inconsistency
itself. By construction we also have
  $f^{A5}_{B7}+f^{A5}_{B8}\leq 1$. Unless $f^{A5}_{B7} = f^{A5}_{B8}
  =0.5$, at least one additional connection is removed.

This selection criteria can thus be too severe as it may
  remove connections that do not lead to inconsistencies.
In this paper we used the normalized shared merit function, which
translate as $\mathcal{M}_m(X,Y) = f^X_Y \times f^Y_X$. As we still
suppose that $M1>M2$, we have $f^{B7}_{A5}\times
f^{A5}_{B7}>f^{B8}_{A5}\times f^{A5}_{B8}$. We still have
$f^{B7}_{A5}>0.5$ and $f^{B8}_{A5}>0.5$ and we know that either
$f^{A5}_{B7}$ or  $f^{A5}_{B8}$ has to be lower than 0.5.
We
could expect that $f^{B8}_{A5}<0.5$ and that the stronger connection
remains. But with $f^{B7}_{A5}$ quite larger than $f^{B8}_{A5}$
($f^{B7}_{A5}\sim 1$ and  $f^{B8}_{A5}\sim 0.5$), we could have
$f^{A5}_{B7}<0.5<f^{A5}_{B8}$ while respecting the ordering of the
merits. The stronger connection (between A5 and B7) would be removed
and only the second best (between A5 and B8) may remain.
A more flexible variation of this criteria would be necessary
 to avoid removing connections one may find important. The conditions
 $f^X_Y \geq 0.5$ and $f^Y_X \geq 0.5$ implies that
 $\mathcal{M}_m(X,Y) \geq 0.25$. This threshold of 0.25 is quite
 efficient in the most majority of case but it doesn't rule out
 either $f^X_Y < 0.5$ or $f^Y_X < 0.5$. This fiducial threshold is not sufficient to avoid
  all possible issues, but too large a selection threshold would strongly limit
the cross-identified catalog.

An alternative approach could be to maximize the number of
cross-identified pairs. Since the cross-comparison is
  structured as a tree, it is straightforward
to detect such problematic configurations as illustrated in \figref{fig:err_tree}. Once the pattern is found, one can simply
ignore the second connection $M2$ and correct the problem automatically
as displayed in \figref{fig:tree_correct}. In the end, we favor a
cross-identification to a simple association.
 The advantage of this
method is that it is easily applied to any kind of merit functions,
whether the criteria is based on distance or on traced mass. The
downside is that we may artificially build
cross-identifications from low strength connections. For example, we could have  $f^{B8}_{A6}<0.1$, which would make
$\mathcal{M}_m(A6,B8) < 0.05$. Applying a threshold on the merit
function thus remains relevant even after inconsistencies have been edited
out. This is the correction we implemented for this
paper. This way we ensure that all inconsistencies are
  avoided independently of any value we may chose for a merit threshold.

\end{appendix}

\end{document}